\documentclass[11pt]{article}

\usepackage{bbm}
\usepackage{float}
\usepackage{amssymb,amsthm,amsmath}
\usepackage{stmaryrd}
\usepackage{eurosym}
\usepackage{color}
\usepackage[a4paper, top=2.5cm, bottom=2.5cm, left=2cm, right=2cm]{geometry}
\usepackage[pdftex]{graphicx}
\usepackage[utf8]{inputenc}
\usepackage{placeins}
\usepackage{bm}
\usepackage{multirow}
\usepackage{mathtools}
\usepackage{authblk}
\usepackage{booktabs}
\usepackage{mbenotes}
\usepackage{algpseudocode}
\usepackage{algorithm}
\usepackage[flushleft]{threeparttable}
\usepackage[toc,page]{appendix}

\usepackage{subfigure}
\usepackage{color}
\usepackage{bbm}
\usepackage{epstopdf}
\usepackage{amssymb,amsthm,amsmath}
\usepackage{stmaryrd}
\usepackage{bm}
\usepackage{multirow}

\theoremstyle{plain}

\theoremstyle{definition}

\theoremstyle{remark}

\algnewcommand{\LineComment}[1]{\State \(//\) \textit{#1}}

\graphicspath{{./figs/}}
\providecommand{\keywords}[1]{\textbf{\textit{\footnotesize Keywords :}} \footnotesize #1}

\DeclareRobustCommand{\officialeuro}{%
  \ifmmode\expandafter\text\fi
  {\fontencoding{U}\fontfamily{eurosym}\selectfont e}}

\begin{document}

\title{Order-book modelling and market making strategies} 

\author[1]{Xiaofei Lu\thanks{xiaofei.lu@centralesupelec.fr}}
\author[1]{Frédéric Abergel\thanks{frederic.abergel@centralesupelec.fr}}
\affil[1]{Chaire de finance quantitative, Laboratoire MICS, CentraleSupélec, Université Paris Saclay}

\maketitle 

\begin{abstract}
Market making is one of the most important aspects of algorithmic trading, and it has been studied quite extensively from a theoretical point of view. The practical implementation of so-called "optimal strategies" however suffers from the failure of most order book models to faithfully reproduce the behaviour of real market participants.

This paper is twofold. First, some important statistical properties of order driven markets are identified, advocating against the use of purely Markovian order book models. Then, market making strategies are designed and their performances are compared, based on simulation as well as backtesting. We find that incorporating some simple non-Markovian features in the limit order book greatly improves the performances of market making strategies in a realistic context.
\end{abstract}

\keywords{limit order books, Markov decision process, market making}
\normalsize

\section{Introduction}

Most modern financial markets are order-driven markets, in which all of the market participants display the price at which they wish to buy or sell a traded security, as well as the desired quantity. This model is widely adopted for stock, futures and option markets, due to its superior transparency. 
%

\begin{figure}[h!]
\begin{center}
\includegraphics[scale=1.0]{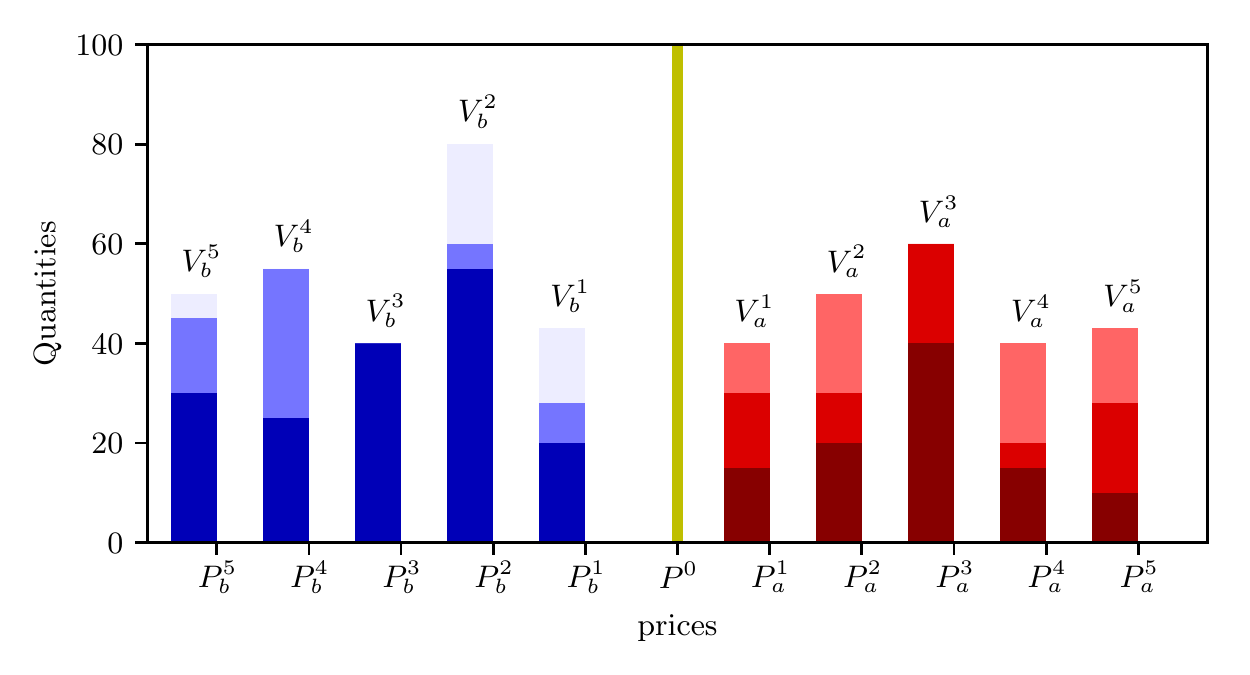}
\caption{\label{Fig:LOB}Illustrative order book example. Blue bars on the left half of the figure represent the available buy orders with prices $P_b^\cdot$ and total quantities $V_b^\cdot$. These correspond to the buyer side, also called the bid side. Participants in the bid side are providing liquidity with prices at which they are ready to buy some quantities of the stock. The right hand bars represent the sell side, commonly called the ask side or offer side, where participants willing to sell post their orders with the prices they are ready to sell the stock. The line in the middle corresponds to the mid-price level and is computed as the average between the best (highest) bid price and the best (lowest) ask price. A transaction occurs when a sell order and a buy order are at least partially matched. A queue of limit orders with the same price is called a limit. Different colours in the same limit represent orders with different priority with darker bars having higher execution priority.}
\end{center}
\end{figure}

In an order-driven market, all the standing buy and sell orders are centralised in the limit order book (LOB). An example LOB is given in Figure \ref{Fig:LOB}, together with some basic definitions. Orders in the LOB are generally prioritized according to price and then to time according to a FIFO rule.


With the emerging of electronic markets, and the deregulation of financial markets, algorithmic trading strategies have become more and more important.

In particular, market making - or: liquidity providing - strategies lay at the core of modern markets. Since there are no more \emph{designated market makers}, every market participant can, and sometimes must, provide liquidity to the market, and the design of optimal market making strategies is a question of crucial practical relevance.

Originating with the seminal paper \cite{Ho:1981}, many researchers in quantitative finance have been interested in a theoretical solution to the market making problem. It has been formalised in \cite{Avell:2008} using a stochastic control framework, and then extended in various contributions such as \cite{Gueant:2013}\cite{Cartea:2013,Cartea:2014,Cartea:2013-2} \cite{Fodra:2013}\cite{Fodra:2015} \cite{Guilbaud:2013}\cite{Guilbaud:2013-2} \cite{bayraktar2014liquidation} \cite{gueant2012optimal} or \cite{gueant2015general}. It is noteworthy that, in this series of papers, the limit order book is not modelled as such, and the limit orders are taken into account indirectly thanks to some probability of execution.

In practice, the price discontinuity and the intrinsic queueing dynamics of the LOB make such simplifications rather simplistic as opposed to real markets, and it is obvious to the practicioner that these actual microstructural properties of the order book play a fundamental role in assessing the profitability of market making strategies. There now exists an abundant literature on order book modelling, and the reader is referred to \cite{abergel2016limit} for an extensive study of the subject, but, as regards market making strategies - or more general trading strategies, for that matter - only very recent papers such as \cite{abergel2017algorithmic} actually address the market making problem using a full order book model.

It is our aim in this paper to contribute to the literature on the subject, both from the modeling and strategy design points of view, so that the paper is twofold: it analyzes and enhances the queue-reactive order book model proposed by \cite{Huang:2015}, and then study the optimal placement of a pair of bid-ask orders as the paradigm of market making.

A word on data: we use the Eurostoxx 50 futures data for June and July, 2016 for the entire analysis, and backtest until Novembre for out-of-sample validation. Eurostoxx 50 futures offers two main advantages:
\begin{enumerate}
 \item it is a very large tick instrument, with an average spread very close to 1 tick and extremely rare multiple-limit trades (less than $0.5\%$);
 \item the value of a futures contract is very high in euros, so that one thinks in terms of number of contracts rather than notional. This actually simplifies the choice of the unit.
\end{enumerate}
These two observations allow us to follow only the first (best) Bid and Ask limits, and focus on the question of interest to us, namely, the design of a model where the state of the order book as well as the type of the order that lead the book into its current state, are relevant. This approach, departing from the purely Markovian case, is based on empirical observations and will be shown to provide a more realistic and useful modelling framework. In a different mathematical setting, a similar reasoning is at the root of Hawkes-process-based order book models such as studied in \cite{lu2017high}\cite{abergel2015long}.


The paper is organized as follows: Section \ref{Queue-reactive discussion} presents the rationale and the calibration of the enriched queue reactive model that improve the performances of the initial model of \cite{Huang:2015}. Section \ref{MMwithMC} addresses the optimal market making strategies in the context of this enhanced model, studying it both in a simulation framework, and in a backtesting engine using real data. 

\section{Challenging the queue-reactive model}
\label{Queue-reactive discussion}
This section presents empirical findings that lay the ground for two improvements to the queue-reactive model of \cite{Huang:2015}. The first one is concerned with the distribution of order sizes, whereas the second, and maybe more original one, addresses the difference in nature of events leading to identical states of the order book.

These improvements will be incorporated in two order book models inspired by, but largely extending, the queue-reactive model. In Section \ref{MMwithMC}, these models will be used in a simulation and backtesting framework to study optimal market making policies. 

\subsection{The queue-reactive model}

In Huang et al. \cite{Huang:2015}, the authors propose an interesting Markovian limit order book model. The limit order book (LOB in short) is seen as a $2K-$ dimensional vector of bid and ask limits $[Q_{-i}:i = 1,\ldots,K]$ and $[Q_{i}:i=1,\ldots,K]$, the limits being placed $i-0.5$ ticks away from a reference price $p_{ref}$.

Denoting the corresponding quantities by $q_i$, the $2K-$dimensional process $X(t) = (q_{-K}(t),\ldots, q_{-1}(t), q_{1}(t), \ldots,q_K(t))$ with values in $\Omega=\mathbb{N}^{2K}$ is modeled as a continuous time Markov chain with infinitesimal generator $\mathcal{Q}$ of the form:
\begin{align*}
 \mathcal{Q}_{q, q+e_i} &= f_i(q),\\
 \mathcal{Q}_{q, q-e_i} &= g_i(q),\\
 \mathcal{Q}_{q, q-e_i} &= -\sum_{q\in \Omega, p\neq q}\mathcal{Q}_{q,p},\\
 \mathcal{Q}_{q, q-e_i} &= 0 \, \text{ otherwise}.
\end{align*}

The authors study several choices for the function $g$: in the first and simplest one, queues are considered independent. The second one introduces some one-sided dependency, whereas the third one emphasizes the interaction between the bid and ask sides of the LOB. Some important statistical features of the limit order book can be reproduced within this model, such as the average shape of the LOB. However, when trying to calibrate the queue-reactive model on our dataset of EUROSTOXX50 future, we observe new phenomena that lead us to enrich the model in two directions.

\subsection{The limitation of unit order size}
\label{Subsection: constant_reference_regime}

\begin{figure}[ht!]
\center
\includegraphics{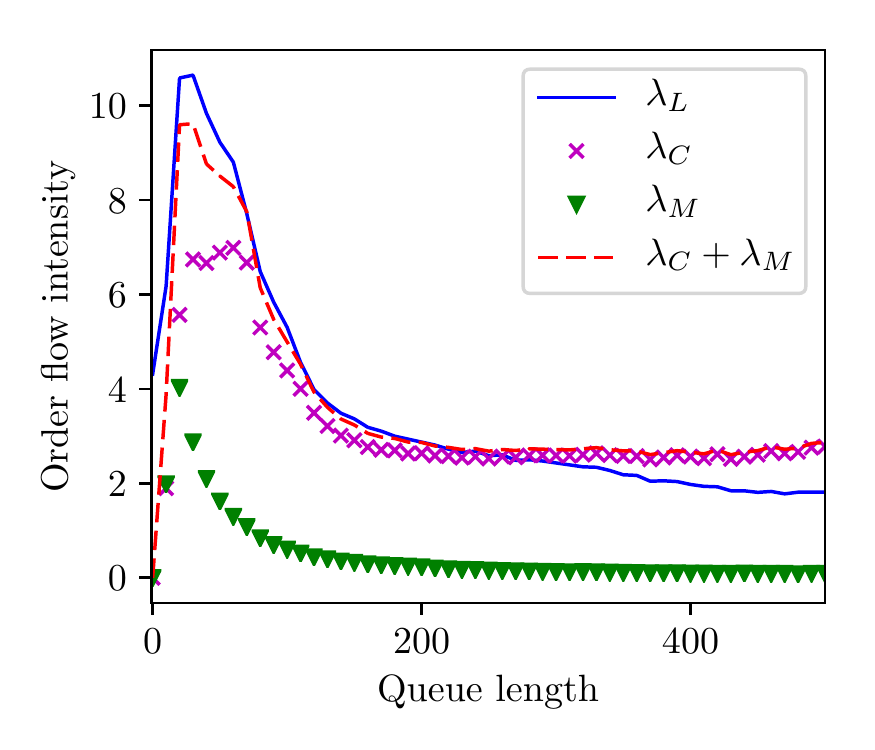}
\caption{Queue reactive model order intensities}
\label{fig: intensity_Huang}
\end{figure}

Following a procedure similar to that in \cite{Huang:2015}, the conditional intensities of limit orders, cancellation and market orders are calibrated and presented in Figure \ref{fig: intensity_Huang}. Note that the intensities of all order types are higher when the corresponding queue length is small. For each queue (bid and ask), the limit orders are liquidity constructive events and the other two are liquidity destructive.

There clearly are three different regimes for the queue sizes:
\begin{itemize}
 \item $\lambda_L$ is slightly higher than $\lambda_C+\lambda_M$ when the queue size is smaller than (approximately) 70.
 \item They become comparable when the queue size lies between 70 and 300.
 \item When the queue size is above 300, $\lambda_L$ decreases whereas $\lambda_C+\lambda_M$ stays stable.
\end{itemize}
Of special importance is the condition $\lambda_L<\lambda_C+\lambda_M$ when the queue size is large, a fact which guarantees that the system is ergodic.

Another interesting feature is that the intensity of market orders drastically decreases when the queue size increases, a fact that can be reformulated as the concentration of trades when the queue size is small. From a practical point of view, a small queue usually indicates a directional consensus, so that liquidity consumers race to take the liquidity before having to place limit orders and wait for execution at the same price.


\begin{figure}[ht!]
\center
\includegraphics{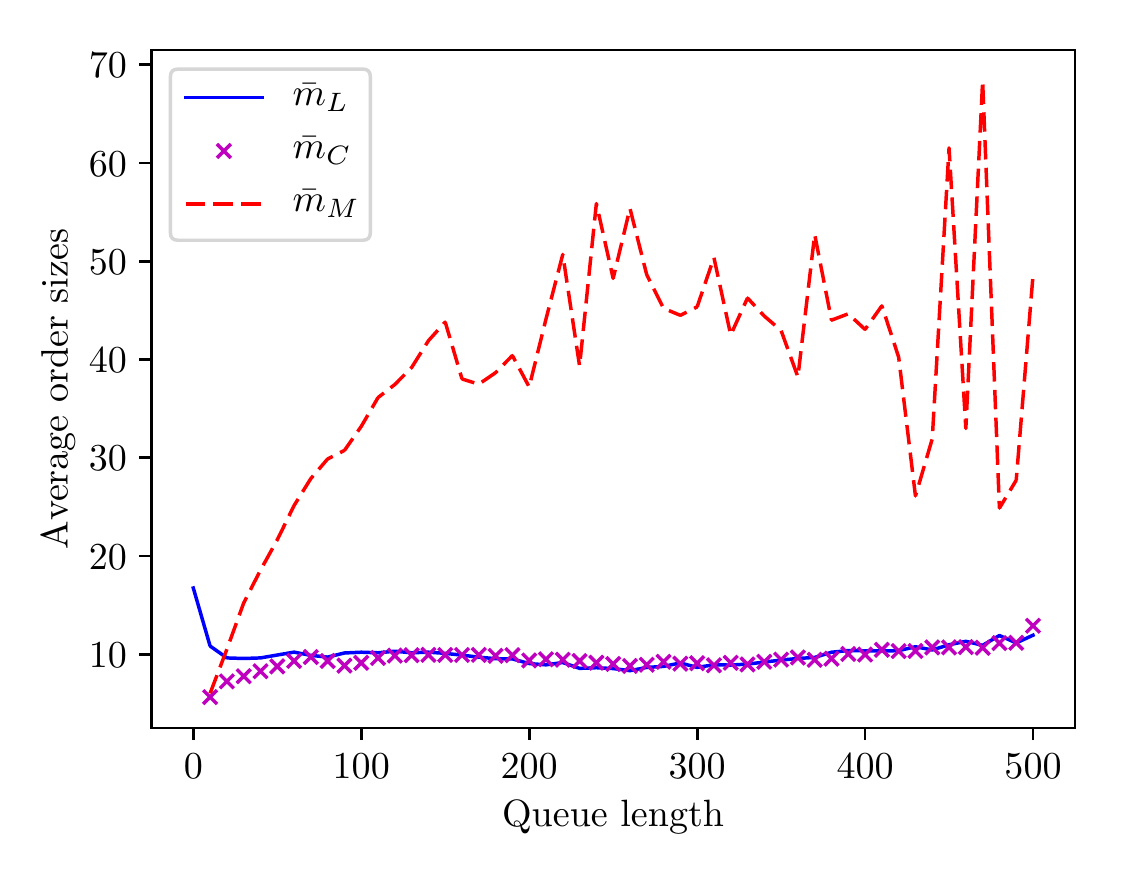}
\caption{Queue reactive model mean order sizes}
\label{fig: mean_size_Huang}
\end{figure}

In the work of \cite{Huang:2015}, and many other related works, the order size is supposed to be constant. It is however clear from empirical analyses that the order sizes have remarkable statistical properties, and that such information is relevant to the LOB dynamics, see for instance \cite{abergel2016limit}\cite{muni2015order}\cite{rambaldi2017role}.

The mean order sizes conditional on the queue states are shown in Figure \ref{fig: mean_size_Huang}. The size of limit and cancellation orders appear to be quite stable across different queue sizes (except for very small queues), whereas the average size of market orders is clearly increasing with the queue size.

\begin{figure}[ht!]
\center
\includegraphics{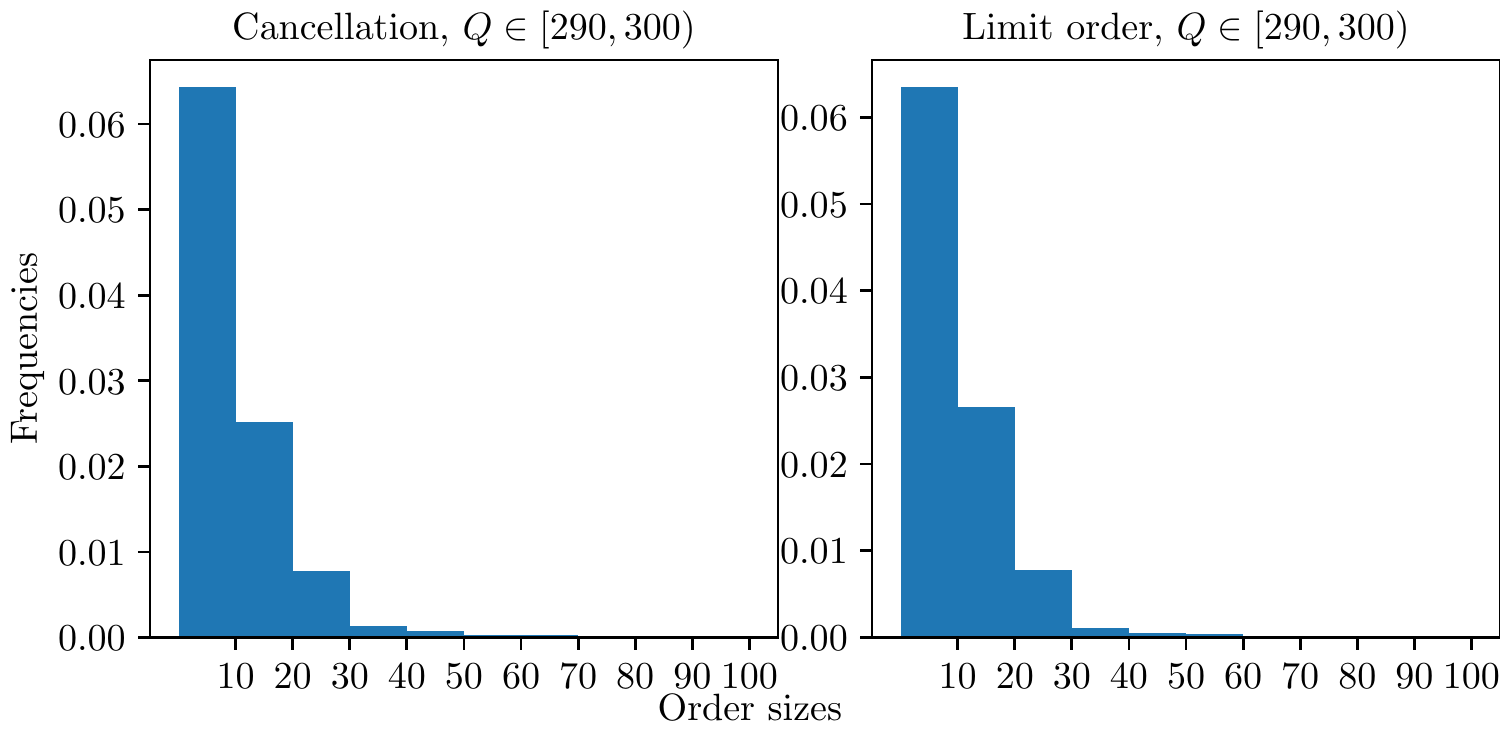}
\caption{Limit order and cancellation size histogram}
\label{fig: lim_cxl_hist}
\end{figure}

\begin{figure}[ht!]
\center
\includegraphics{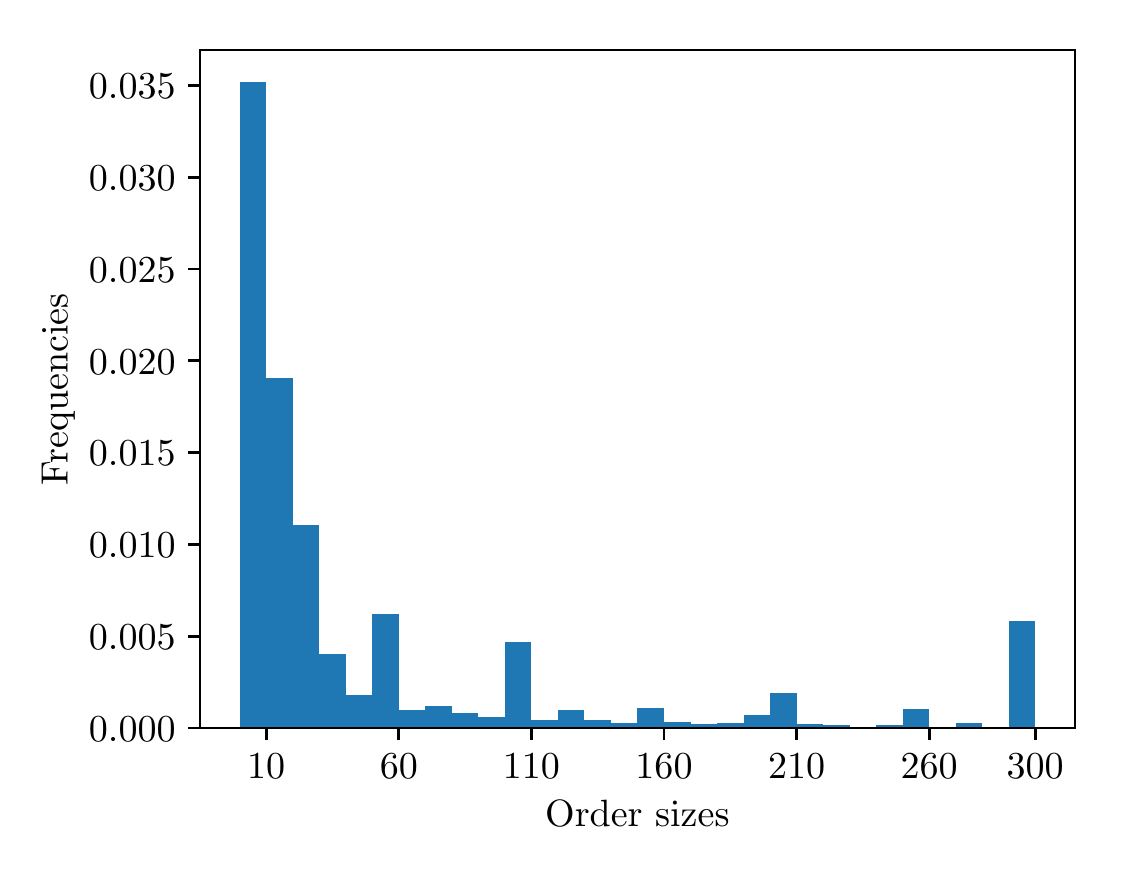}
\caption{Market order size histogram when $Q\in[290,299)$}
\label{fig: mkt_hist}
\end{figure}

To further analyze the distributions of order sizes, histograms of order sizes with the queue length in the interval of $[2900,299)$ are shown in Figure \ref{fig: lim_cxl_hist} and \ref{fig: mkt_hist}. Bins of 10 futures on the x-axis are used to produce the histograms. The first bin 

The empirical distributions for limit orders and cancellations are rather similar and can be modelled with a geometric distribution. On the contrary, there exist some interesting patterns in the sizes of market orders: the distribution looks like a mixture of a geometric distribution for small sizes, and Dirac functions at $\{[50,59), [100,109), [150,159), \cdots\}$ and $q=Q$. It is not surprising, because traders do not necessarily randomize their market orders, so that multiples of 50 occur quite frequently, and so do orders that completely eat up the first limit. For the rest of this subsection, we denote $q, Q \in\mathbb{N}^*$ to represent the bin of $[10(q-1), 10q)$ and $[10(Q-1), 10Q)$.

Inspired by such observations, a simple model of order sizes can be proposed: 
\begin{itemize}
 \item Limit order sizes follow geometric laws $p_L(q;Q)$ with parameters $p_0^L(Q)$ depending on the queue sizes;
 \item Cancellation sizes follow truncated geometric laws with parameters $p_C(q;Q)$
 $$p_C(q;Q) = \mathbb{P}[q|Q] = \frac{p_0^C(1-p_0^C)^{q-1}}{1-(1-p_0^C)^Q}\mathbbm{1}_{\{q\leq Q\}}$$;
 \item Market order sizes follow a mixture of geometric laws and Dirac functions
 $$p_M(q;Q) = \mathbb{P}[q|Q] = \theta_0\frac{p_0^M(1-p_0^M)^{q-1}}{1-(1-p_0^M)^Q}\mathbbm{1}_{\{ q\leq Q\}}+ \sum_{k=1}^{\lfloor \frac{Q-1}{5} \rfloor} \theta_k\mathbbm{1}_{\{q=5k+1\}}+\theta_\infty \mathbbm{1}_{\{q=Q, Q \neq 5n+1\}}$$
 where the parameters $\{p_0^M, \theta_0, \theta_k, \theta_\infty\}$ depend on $Q$.
\end{itemize}
The parameters of limit order sizes are simply estimated. For the cancellation and market orders, a maximum likelihood method can be used. The market order log-likelihood is
\begin{align*}
 \log (L) &= \sum_{q_i\neq 5n+1, q_i\neq Q}\log\theta_0+(q_i-1)\log(1-p_0^M)+\log p_0^M-\log(1-(1-p_0^M)^Q)\\
 &+\sum_{k=1}^{\lfloor \frac{Q-1}{5} \rfloor} \sum_{q_i=5k+1} \log(\theta_0(1-p_0^M)^{q_i-1}p_0^M+\theta_k(1-(1-p_0^M)^Q))-\log(1-(1-p_0^M)^Q)\\
 &+\mathbbm{1}_{\{Q \neq 5n+1\}}\sum_{q_i=Q} \log(\theta_0(1-p_0^M)^{q_i-1}p_0^M+\theta_\infty(1-(1-p_0^M)^Q))-\log(1-(1-p_0^M)^Q).
\end{align*}

The calibration results are presented in Table \ref{Tab:lim_cxl_size} and \ref{Tab:mkt_size}. As expected, the geometric distribution parameters for limit orders and cancellations are quite close to each other, and remain stable across different queue lengths. The geometric distribution parameter for market orders slightly decreases with the queue length. Dirac parameters $\theta_k$ are very stable across different queue lengths.

\begin{table}[ht!]
\begin{center}
\begin{threeparttable}
\renewcommand{\arraystretch}{1.15}
\caption{Calibrated limit order and cancellation size parameters}
\label{Tab:lim_cxl_size}

\begin{tabular}{c|cccccccccc}
\hline
$Q$  & 21     & 22     & 23     & 24     & 25     & 26     & 27     & 28     & 29     & 30     \\\hline
$p_0^L$ & 0.6421 & 0.6415 & 0.6458 & 0.6410 & 0.6443 & 0.6430 & 0.6418 & 0.6439 & 0.6404 & 0.6387 \\
$p_0^C$ & 0.6578 & 0.6591 & 0.6600 & 0.6623 & 0.6598 & 0.6611 & 0.6557 & 0.6554 & 0.6538 & 0.6496 \\ \hline
\end{tabular}

\end{threeparttable}
\end{center}
\end{table}

\begin{table}[ht!]
\begin{center}
\begin{threeparttable}
\renewcommand{\arraystretch}{1.15}
\caption{Calibrated market order size parameters}
\label{Tab:mkt_size}
\begin{tabular}{ccccccccccccc}
\hline
$Q$ & $p_0^M$ & $\theta_0$ & $\theta_1$ & $\theta_2$ & $\theta_3$ & $\theta_4$ & $\theta_5$\tnote{*} & $\theta_6$\tnote{*} \\\hline

21 & 0.3486 & 0.8357 & 0.0185 & 0.0338 & 0.0081 & 0.1038 & - & -\\
22 & 0.3557 & 0.8311 & 0.0198 & 0.0338 & 0.0094 & 0.0215 & 0.0844 & -\\
23 & 0.3383 & 0.8517 & 0.0148 & 0.0366 & 0.0084 & 0.0188 & 0.0697 & -\\
24 & 0.3327 & 0.8475 & 0.0108 & 0.0373 & 0.0099 & 0.0192 & 0.0753 & -\\
25 & 0.3333 & 0.8310 & 0.0234 & 0.0379 & 0.0084 & 0.0214 & 0.0779 & -\\
26 & 0.3292 & 0.8391 & 0.0203 & 0.0408 & 0.0115 & 0.0167 & 0.0716 & -\\
27 & 0.3250 & 0.8374 & 0.0188 & 0.0369 & 0.0114 & 0.0220 & 0.0116 & 0.0619 \\
28 & 0.3134 & 0.8351 & 0.0191 & 0.0452 & 0.0086 & 0.0201 & 0.0164 & 0.0554 \\
29 & 0.3090 & 0.8262 & 0.0192 & 0.0476 & 0.0086 & 0.0181 & 0.0135 & 0.0668 \\
30 & 0.3050 & 0.8426 & 0.0205 & 0.0402 & 0.0096 & 0.0188 & 0.0103 & 0.0580 \\ \hline

\end{tabular}

\begin{tablenotes}
\item[*] For the cases of $Q \neq 5k+1$, $\theta_{\lfloor \frac{Q-1}{5}\rfloor+1}$ represents $\theta_\infty$
\end{tablenotes}

\end{threeparttable}
\end{center}
\end{table}

The enhanced LOB model is then driven by compound Poisson processes with intensities conditional on the queue size.

\subsection{The role of limit removal orders}

\label{Subsection: limit_removal_order}

By construction, Markovian LOB models assume that the past has no influence on the future except through the present. Nevertheless, it has been established, see e.g. \cite{abergel2016limit}\cite{lu2017high} for some in-depth empirical studies, that the nature of past events actually influence the order flow and, therefore, the future states of the LOB.

In this section, the emphasis is set on the nature of the last event that totally removes the liquidity at one limit, and the subsequent evolution of the LOB.

An order that completely eats up a limit is termed a \emph{limit removal order}. Such an order can only be a cancellation or market order. The liquidity removal process is denoted by $Y(t) = (q^r(t), O^r(t))$, where $q^r$ stands for the order size and $O^r\in\{O^M, O^C\}$ for its type - $O^M$ for a market order and $O^C$ for a cancellation. Then, $Y(t)$ is a càdlàg process with jump times $\tau^r = \{\tau^r_i\}_{\{i=1,2,\ldots\}}$.

In a symmetric way, an order that creates a new limit will be termed a \emph{limit establishing order}. Such an order can only be a limit order, but it can be placed either on the bid or ask side since the limit is empty. The liquidity establishing order process is denoted by $Z(t) = (q^e(t), O^e(t))$, where $O^e\in\{F, R\}$: $F$ for \lq follow \rq, so that $P(\tau^e_i)\neq P(\tau^r_i-)$, and $R$ for \lq revert \rq). Again, $Z(t)$ is a càdlàg process with jump times as $\tau^e = \{\tau^e_i\}_{\{i=1,2,\ldots\}}$.
%
%

\subsubsection{Characterization of $O^e$}
\label{Subsubsection: Ne_characterization}

\begin{table}[ht!]
\begin{center}
\begin{threeparttable}
\renewcommand{\arraystretch}{1.15}
\caption{$O^e$ conditional probabilities}
\label{Tab: events_type_count}

\begin{tabular}{c|cc|c}
\hline
$O^r$  &   $F$   &  $R$ & $\mathbb{P}[O^e=F| O^r]$\\\hline
 $O^M$& 7554 & 1409 & $84.3\%$\\
 $O^C$& 1043 & 2823 & $ 27.0\%$\\ 
\hline

\end{tabular}

\end{threeparttable}
\end{center}
\end{table}

We first investigate $O^e(\tau^e_i)$ conditional on $O^r(\tau^r_i)$. Table \ref{Tab: events_type_count} presents the daily average number of events as well as their conditional probabilities.

A first important observation is that the price tends to move in the same direction if it is triggered by $O^M$, whereas mean-reversion is more likely in the case of an $O^C$-triggered price change. In other words, $O^M$ is more informative than $O^C$, and is the main driver of price moves.

\begin{figure}[ht!]
\center
\includegraphics{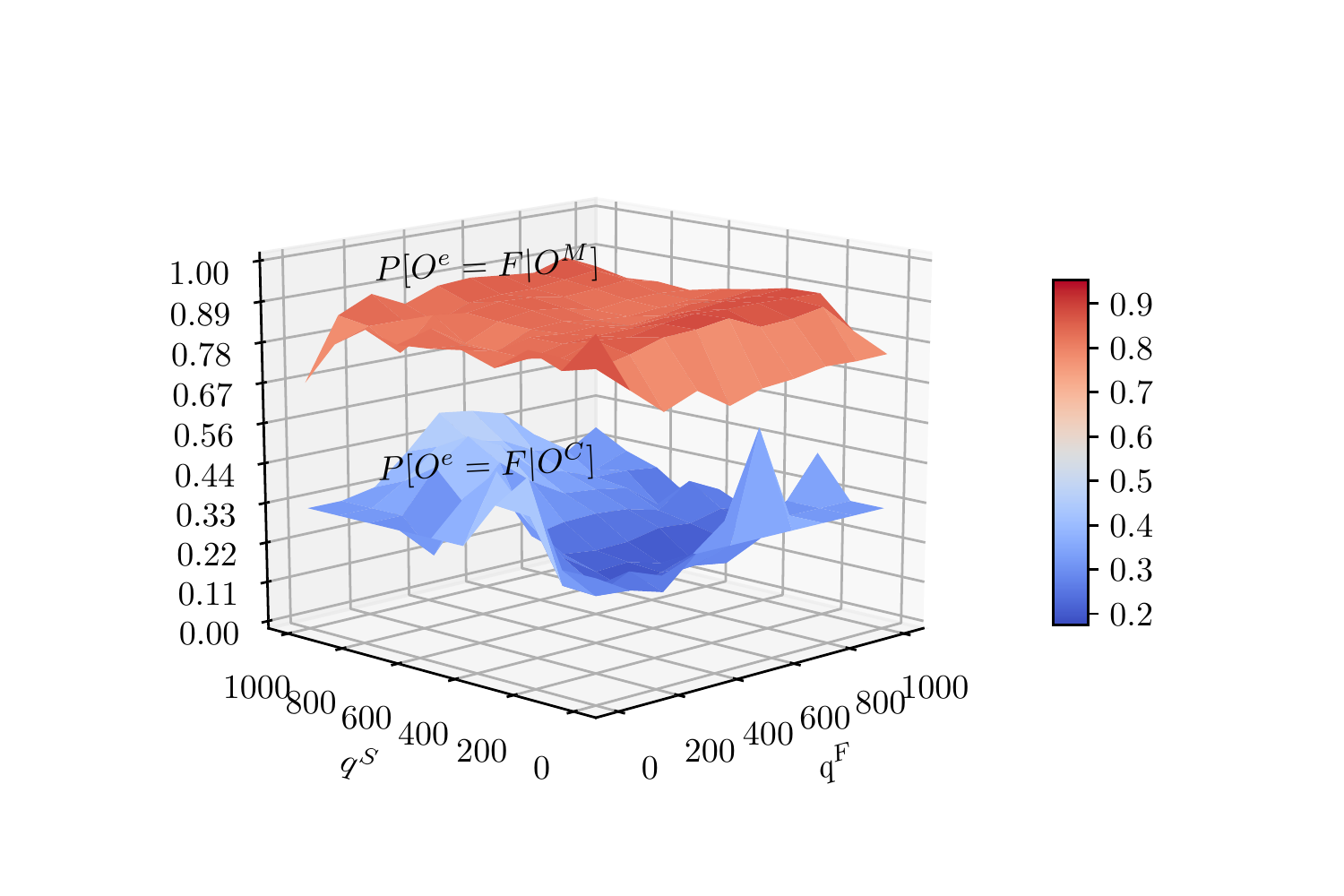}
\caption{Conditional distribution $O^e$ w.r.t the first limit queue sizes for different $O^r$}.
\label{fig: independence_test}
\end{figure}

One can wonder whether this dependency structure could be simplified to one on the state of the LOB only, that is, whether the conditional distribution of $O^e|O^r, X$ could be explained by $O^e|X$. Figure \ref{fig: independence_test} presents the empirical distributions of $O^e|(q^F, q^S)$ for $O^r \in \{O^M,O^C\}$ respectively, where $q^F = q_{-1}$ and $q^S = q_2$ at $\tau^e_i-$, and $q^F = q_1$ and $q^S = q_{-2}$ for the corresponding bid side, which are LOB states just before the arrival of a limit establishing event.

Although the queue sizes may vary before the arrival of a limit establishing event, Figure \ref{fig: independence_test} shows that their influence is negligible, and the conclusion is that $O^e$ is highly dependent on the $O^r$ but much less on $(q^F, q^S)$. 

\begin{figure}[ht!]
\center
\includegraphics{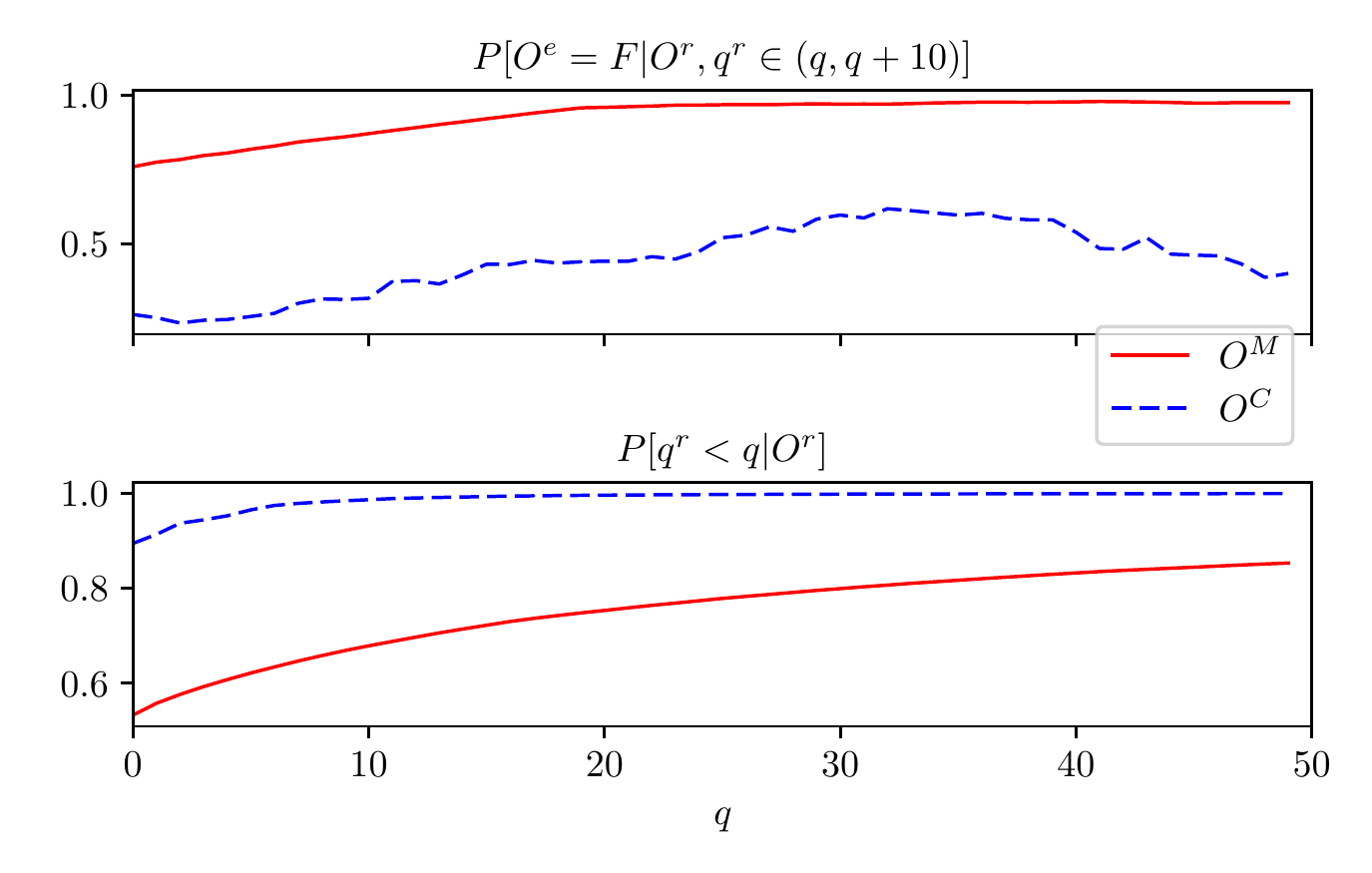}
\caption{Conditional distribution $O^e$ w.r.t $q^r$ for different $O^r$}.
\label{fig: spread_switch_qty2}
\end{figure}

It seems also relevant to include the sizes $q^r(\tau^r_i)$ in the analysis. In Figure \ref{fig: spread_switch_qty2}, the upper panel presents the conditional distribution of $O^e|O^r, q^r$ as a function  of $q^r$, while the lower panel presents the cumulative distribution function of $q^r$. For both $O^M$ and $O^C$, the probability of $O^e=F$ increases with the order size. It is however noteworthy that, for $O^M$, this probability reaches almost 1 when $q^r\geq 30$, a rather significant fact as there are over $20\%$ of market orders that lay in the interval of $(30,+\infty)$.

For $O^C$, the situation is completely different: not only does the probability of $O^e=F$ remains close to 0.5, but the proportion of cancellation orders of size larger than 20 is very small and the decrease of $\mathbb{P}[O^e=F|O^r=O^c, q^r]$ when $q^r>50$ is not statistically significant.


The interpretation of this phenomenon is direct: not only market orders are much more informative than cancellations but, the larger the market order is, the more likely it is to indicate a directional price movement that the market will follow. A market buy (sell) order of size larger than 30 that consumes the entire liquidity will almost always be followed by new liquidity providers placing bid (ask) limit orders at the previous trade price.

\subsubsection{The size of liquidity establishing orders}

Table \ref{Tab: events_qty_count} summarizes the average number of $(O^r, O^e, q^r)$ events, as well as the average quantity $q^e$ corresponding to each of these classes of events. As already observed in Section \ref{Subsubsection: Ne_characterization}, very few large cancellations result in an empty limit, most large orders are market orders that systematically lead to a follow event. Moreover, one can clearly see a monotonically increasing relationship between $q^e$ and $q^r$ across all values of $(O^r, O^e)$.

Due to the low number of large orders in other classes, we will concentrate on $(O^M, F)$ for a in-depth analysis.

\begin{table}[ht!]
\begin{center}
\begin{threeparttable}
\renewcommand{\arraystretch}{1.15}
\caption{$q_e$ conditional distribution on $O^r$, $O^e$ and intervals of $q^r$}
\label{Tab: events_qty_count}

\begin{tabular}{c|ccc|ccc}
\hline
&\multicolumn{3}{c|}{\# of events per day} & \multicolumn{3}{c}{mean $q^e$ size}\\\hline

$(O^r, O^e)$& $(0, 10)$ &$[10,20)$ &$[20, \infty)$ & $(0, 10)$ &$[10,20)$ &$[20, \infty)$ \\\hline
$(O^M, F)$& 3623 & 1128 & 2802 & 7.46, & 10.18 & 27.93 \\
$(O^M, R)$& 1153 & 180 & 75 & 6.69, & 6.75 & 27.98 \\
$(O^C, F)$& 906 & 112 & 23 & 7.22, & 7.45 & 9.08 \\
$(O^C, R)$& 2552 & 243 & 26 & 5.60, & 7.71 & 7.99 \\
\hline
\end{tabular}
\end{threeparttable}
\end{center}
\end{table}

\begin{figure}[ht!]
\center
\includegraphics{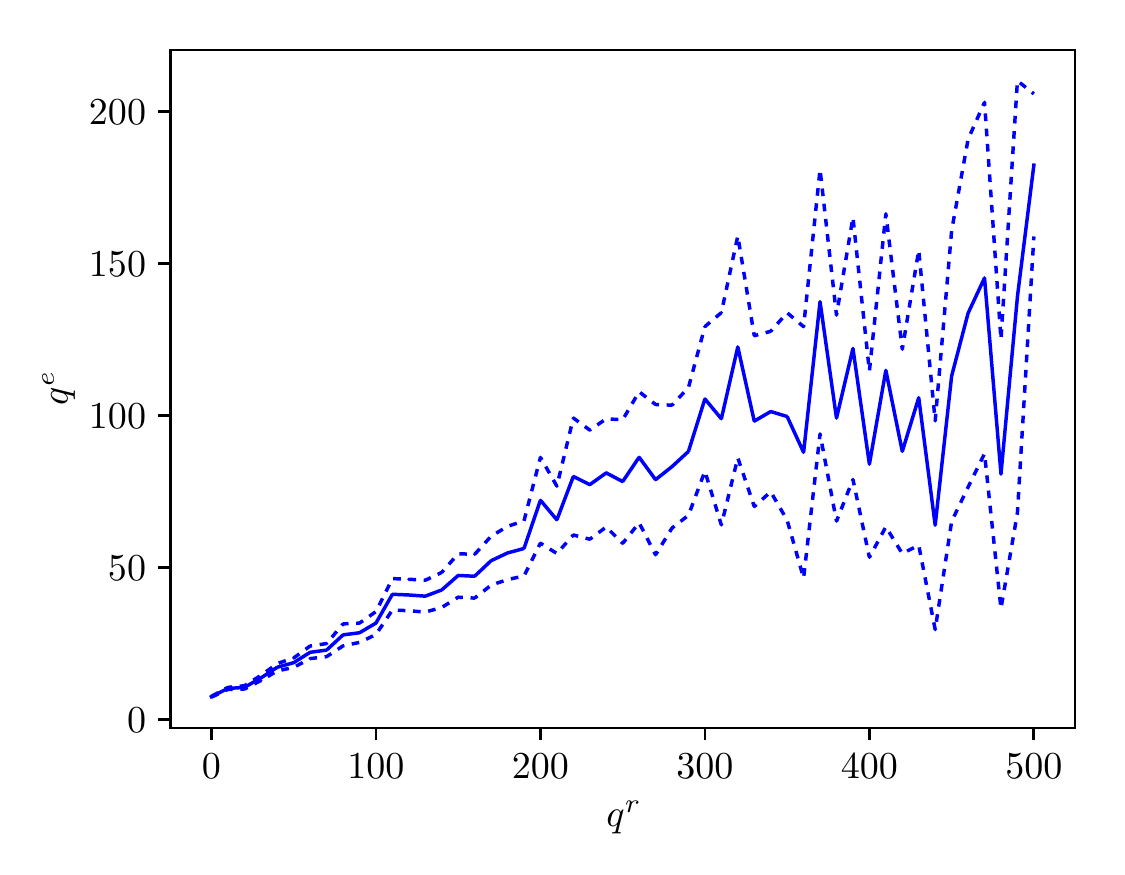}
\caption{Mean $q^e$ with respect to $q^r$. The full line presents the mean and the dashed line presents the $95\%$ confidence interval of the estimation.}
\label{fig: spread_switch_qe}
\end{figure}

Figure \ref{fig: spread_switch_qe} illustrates the variation of $q^e$ with respect to $q^r$ for $(O^r, O^e) = (O^M, F)$. There exists a definite monotonically increasing relationship between the two quantities, and the new limit can even reach a size of over 100 shortly after the old limit is consumed by a very large market order. Clearly, the geometric distribution that we have previously  advocated for the size of limit orders fails to represent such a phenomenon.

\begin{figure}[ht!]
\center
\includegraphics{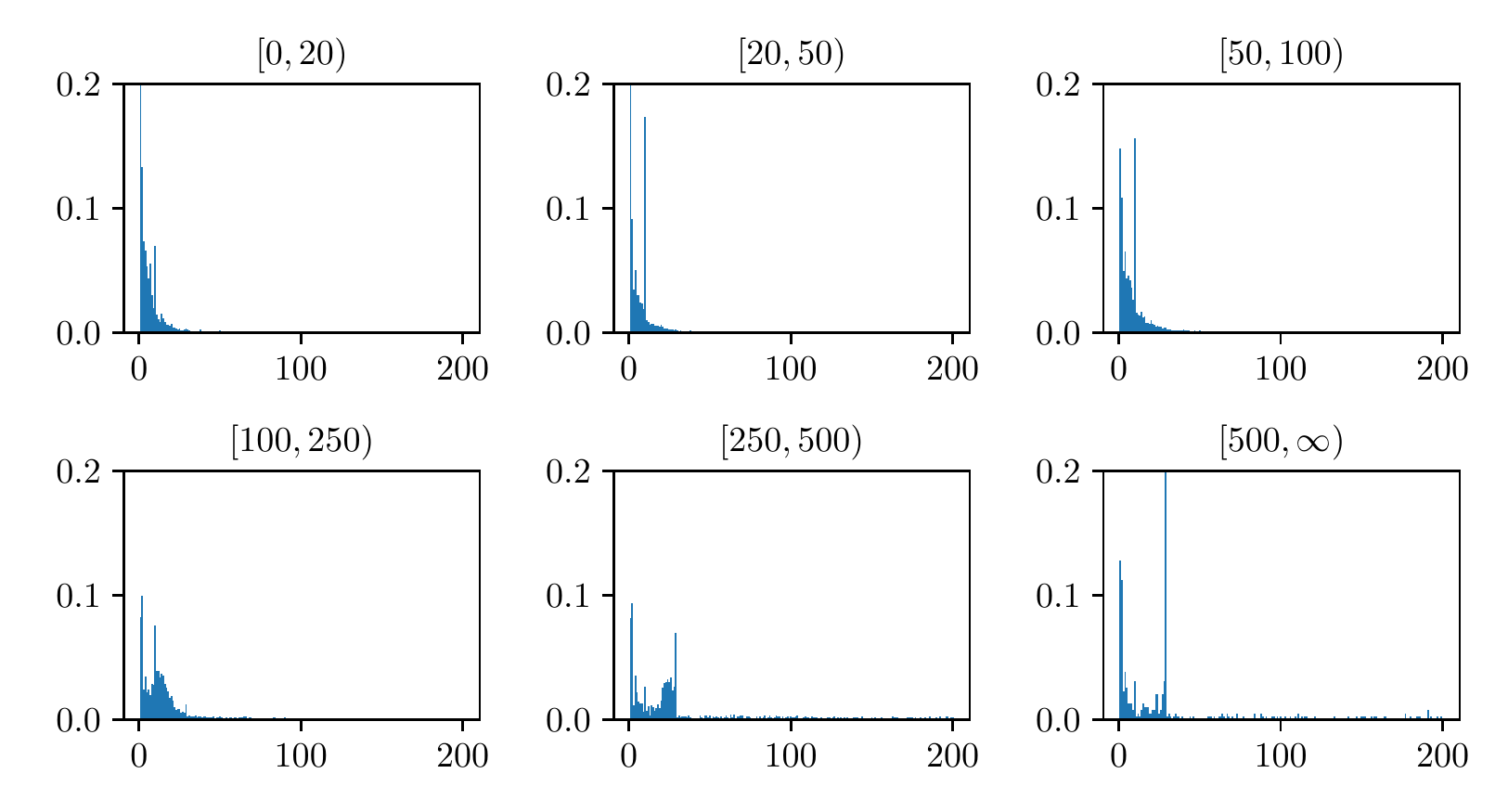}
\caption{Histograms of $q^e$ with respect to $q^r$ in different intervals.}
\label{fig: spread_switch_qe1}
\end{figure}

Figure \ref{fig: spread_switch_qe1} shows some empirical conditional distributions of $q^e$ obtained by classifying $q^r$. When $q^r$ is small, the distribution is close to geometric. However, as $q^r$ increases, the density presents fatter tails and discrete peaks, and it is no longer appropriate to use a geometric distribution - one may rather consider using the empirical distribution to model $q^e|q^r$.

\subsubsection{Characterizing inter-event durations}

\begin{figure}[ht!]
\center
\includegraphics{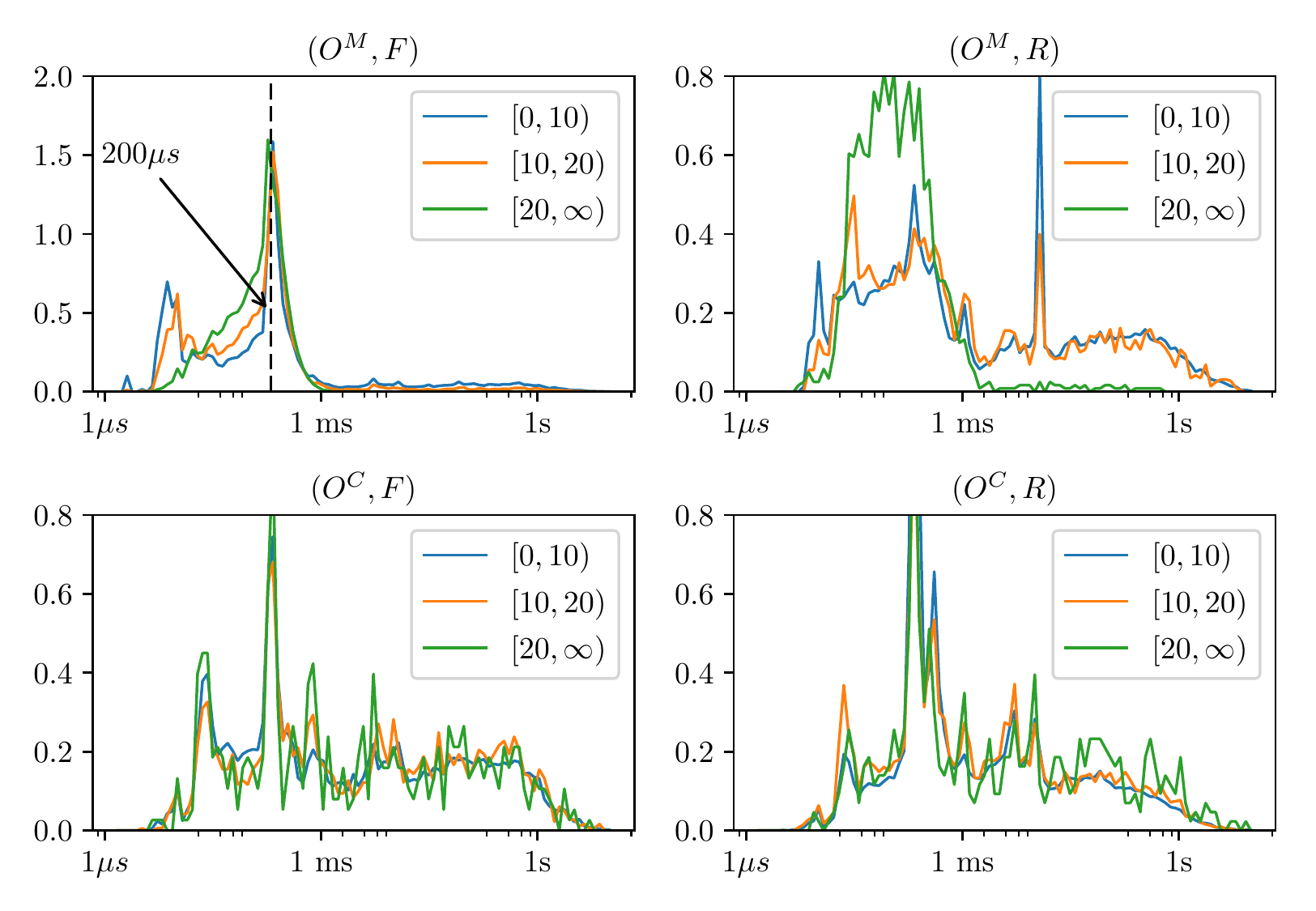}
\caption{$\log_{10}(\tau^r-\tau^e)$ distribution with respect to $O^r$, $O^e$ and $q^r$ in different intervals.}
\label{fig: spread_switch_dt}
\end{figure}

Going further, we now analyze the empirical distributions of (the logarithm of) inter-arrival times $\log_{10}(\Delta \tau)$ in Figure \ref{fig: spread_switch_dt}. When the triggering event is $O^C$, the distribution of $\Delta\tau$ are very similar, no matter what $O^e$ and the size of the order are. When the triggering event is $O^M$ and the price reverts, $\Delta \tau$ distributions are quite similar when $q^r<20$.

On the contrary, the distribution of $\Delta \tau|(O^M,F)$ is very different from the others, and is different for different intervals of $q^r$. First, as expected, except for very small $q^r$, the densities of $\Delta\tau>1ms$ are close to 0. In addition, there is a sharp peak around $200\mu s$. One possible reason of such concentration is about the reaction to large trades. Though liquidity providers try to place limit orders immediately after a large trade, they are constrained by the round-trip latencies of the market and their own systems. As $200 \mu s$ represents the typical market round-trip latency, a high density concentrates around this level. But there could have been a few market makers who react to exogenous information that is the same as the large trade but arrives late which result in the density below. And the above part comes simply from the higher latencies of some market makers.

\subsection{Enriching the queue-reactive model}
\label{Models}
As a conclusion to this empirical study, one can see that the dynamics of the order book, and not only its state, must be used in order to gain a faithfull representation of the market. In order to enhance the queue-reactive model, it is necessary to add a dependency of the order sizes and, more importantly, a dependency on the nature of the order that drove the book into its current state. One must therefore depart from the Markovian framework, but only slightly, and in the interest of a much more realistic modelling.

We then propose two LOB models that can be viewed as extensions of the queue-reactive model, but differ from it as regards price transitions:
\begin{itemize}
 \item \textbf{Model I}: When either limit is empty, the next limit order that closes the spread depends only on whether the emptied limit was on the bid or ask side. The size of the order and the recurrence time are independent.
 \item \textbf{Model II}: The new limit order $(O^e, q^e)$ depends not only on the side of the cleared limit, but also are functions of the last removal event $(O^r, q^r)$. The arrival time of the event is determined by $\tau^e = \tau^r+\Delta\tau$, where $\Delta\tau$ is dependent on $(O^r, q^r)$.
\end{itemize}

Each of these models is compared with the unit size model having the same transition rule as Model I, which we refer to as Model 0. Monte Carlo simulations are conducted for the different models, and the results are benchmarked against real data.

\begin{figure}[ht!]
\center
\includegraphics{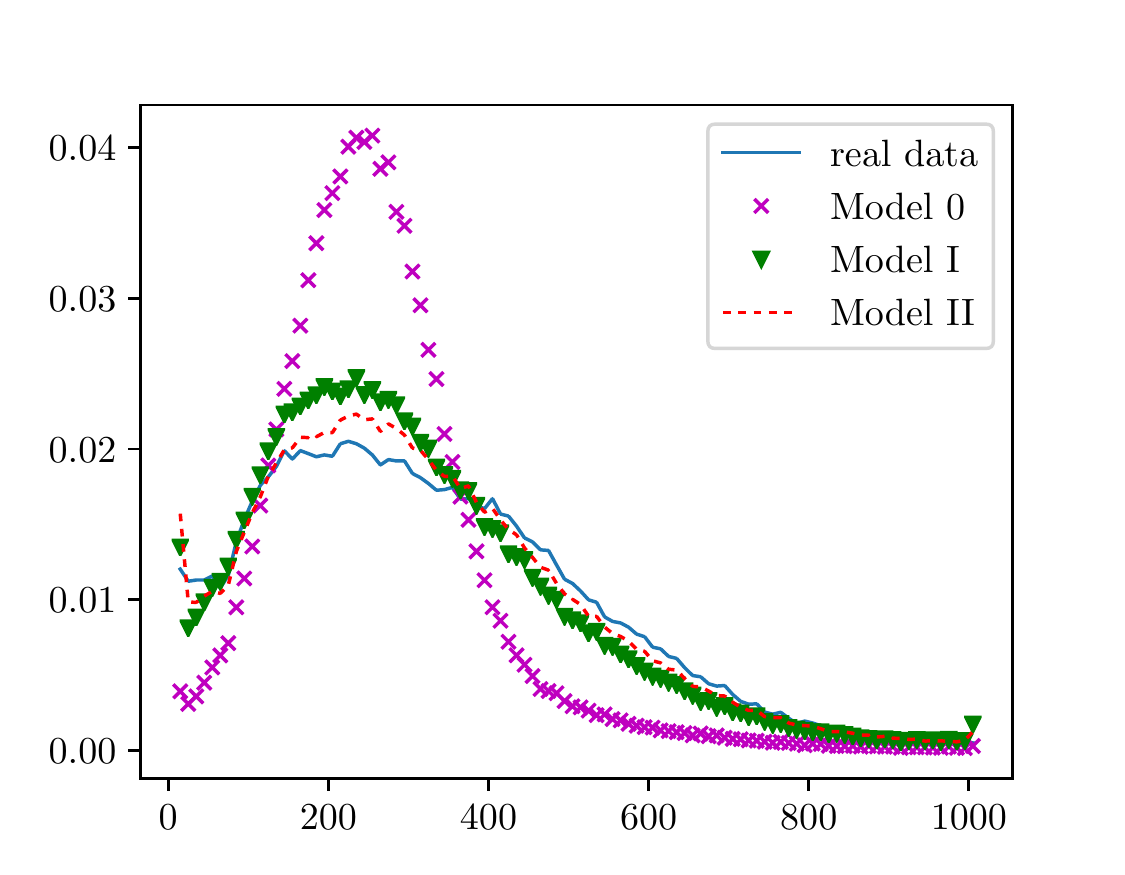}
\caption{Best bid/ask quantities distributions.}
\label{fig: simu_1lim_law}
\end{figure}

As an example, Figure \ref{fig: simu_1lim_law} presents the distributions of the best limit quantities sampled at $1s$ frequency in the various models, as well as in the data. Model 0 produces a LOB that is very concentrated around the constructive-destructive equilibrium limit size (around 300), with lower density for smaller queues and almost no density for long queues. Adding the size distribution in Model I already improves the model, with a global shape closer to the real data. Model II improves the too high density around the equilibrium queue size, and also produces more realistic, fatter tails for the queue size distribution.

\section{Market making in real markets}
\label{MMwithMC}

In this section, we address the problem of defining an optimal market making (or: liquidity providing) policy in an order-driven financial market. Our goal is to mimick as well as possible the situation of an electronic market maker, and a realistic order book model incorporating the various empirical properties we have just shown must therefore be designed. Once a model is set up, stochastic control is used in order to derive the optimal strategy. The characteristics of the optimal strategy are quite useful in determining whether a model makes sense or not, as its performances in a real market environment can be measured, either through backtesting or direct experimentation. An \lq optimal \rq strategy that would lose money in the market would be an indication of poor modelling !

Based on the work in \cite{hult2010algorithmic} (see also \cite{abergel2017algorithmic} \cite{bauerle2011markov} for related approach and results), optimal market making strategies are numerically and empirically studied under the various hypotheses corresponding to the models labelled 0 and II introduced in Section \ref{Models}. In the interest of readability, the theoretical framework and main results are recalled in Appendix \ref{appendix:MDP}, while the current section is devoted to the presentation and discussion of the results.


As usual in stochastic control, the quantity of interest is the \emph{value function} in each state of the LOB, that is, the expected future profit and loss (P\&L) of a pair of bid and ask orders: the risk-neutral market maker (MM) will place an order if the value is strictly greater than 0, and stay out of the market otherwise. The associated optimal strategy is thus defined according to the value functions.

After a simulation-based study and analysis of optimal market making strategies, with or without inventory control, the optimal strategies for each models are backtested against realistic market conditions.

Note that, for technical reasons, the maximum queue size is set equal to 500 contracts and the bid-ask spread is supposed to be always 1 tick - as a matter of fact, for liquid, large-tick instruments, trades almost never occur when the spread is larger than 1 tick, so that this hypothesis stands.







\subsection{Optimal market making strategies}

This section is devoted to a comparison of the optimal strategies for Model 0 and Model II. The results for Model I are only slightly different from Model 0 and will not be presented here.

The state of the LOB is described by a quadruple $(x^B,x^A,y^B,y^A)$ representing the best bid and ask quantities in number of lots, and the position of the MM's orders in the queue. The key question in market making is the time to enter the order book.

Initially, the MM places his order at the end of the queue, so that
$$x^B = y^B \qquad x^A = y^A.$$

\subsubsection{Value function and optimal strategies for Model 0}

\begin{figure}[ht!]
\center
\includegraphics{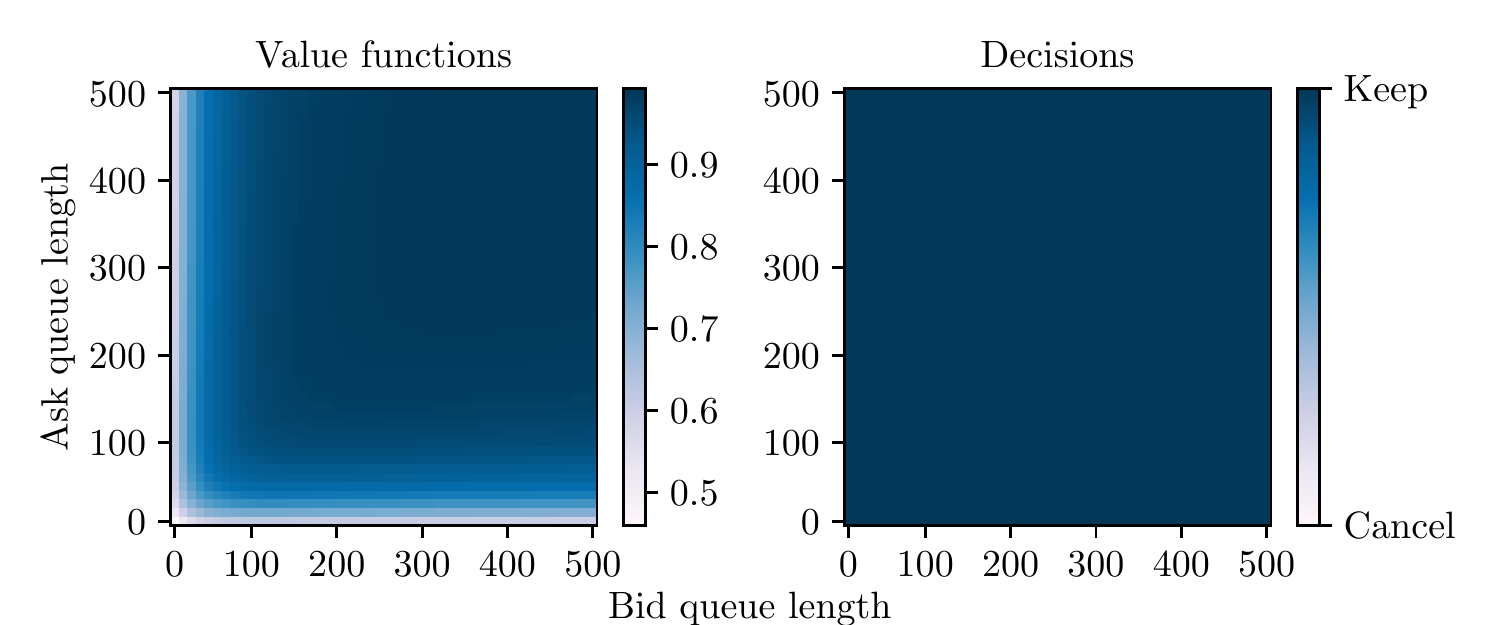}
\caption{Initial state values in Model 0}
\label{fig: values_model0}
\end{figure}

We illustrate the value as a function of the initial state in Figure \ref{fig: values_model0}. The x and y axis are respectively the bid and ask queue lengths in the figure. Once the state values are known, the optimal strategy is staighforward: if the value is positive, the optimal action is to stay in the order book, whereas if the value is 0, the optimal action is to cancel the orders.


One clearly observes that the values are mostly positive. When the order book is highly imbalanced, the order value is lower. The lowest values appear when both of the limit queues are short. However, the minimum value is still higher than 0.5. The result can be interpreted as follows: when the queue length is short, the probability that the queue will become empty increases before other limit orders arrive behind the MM's orders; when his order is executed on one side of the order book but not on the opposite side, the MM will have to close the position at a loss with a market order. When the queue is long, the probability that both MM's orders get executed before the price changes is higher, so the value is closer to 1.

However, such a result is hardly a reflection of reality. From practical experience we know that a simple market making strategy is unprofitable or poorly profitable in the real market. The value functions calibrated with Model 0 are at least 0.5 tick, indicating that we should follow a naive strategy, always placing an order on both the bid and ask sides. This unrealistic behaviour comes from the fact that Model 0 has very stable limits, so that the MM's bid and ask orders tend to be both executed before either limit is cleared.

\subsubsection{Value function and optimal strategies for Model II}

\begin{figure}[ht!]
\center
\includegraphics{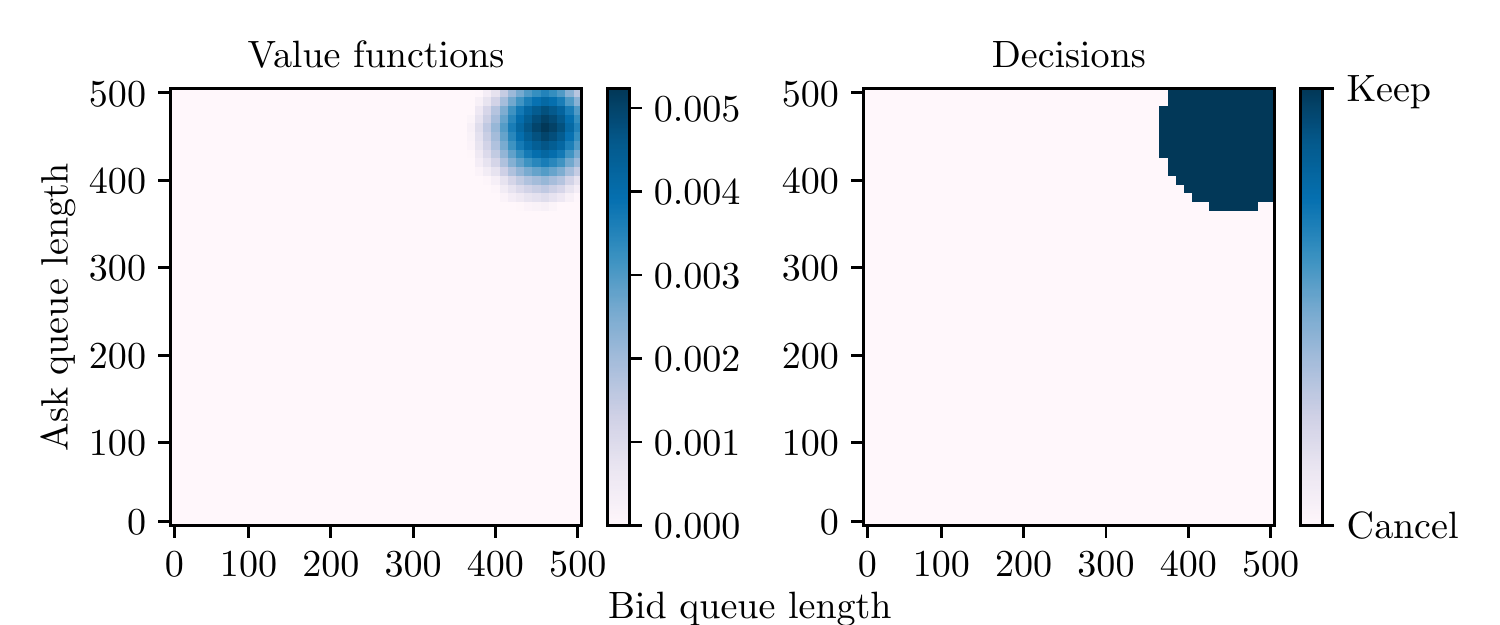}
\caption{Initial state values in enriched Queue-reactive model}
\label{fig: values_Huang}
\end{figure}

Figure \ref{fig: values_Huang} shows the values (left) and the decisions (right) depending on the initial state. The state is non-profitable when the state value is 0, so that the MM cancels orders on both sides and wait for a further transition, otherwise, he stays in the book and wait for his orders to get executed.

Clearly, the results are very different from those obtained previously: in most of the initial states, the optimal policy is to cancel. In fact, the MM is penalized when there is a price change before both his orders get executed, and the existence of large market orders increases such a risk. The most favourable situation is when the two queues are balanced and relatively long, so that his orders can gradually gain priority before the order book becomes imbalanced again and the price changes.

\begin{figure}[ht!]
\center
\includegraphics{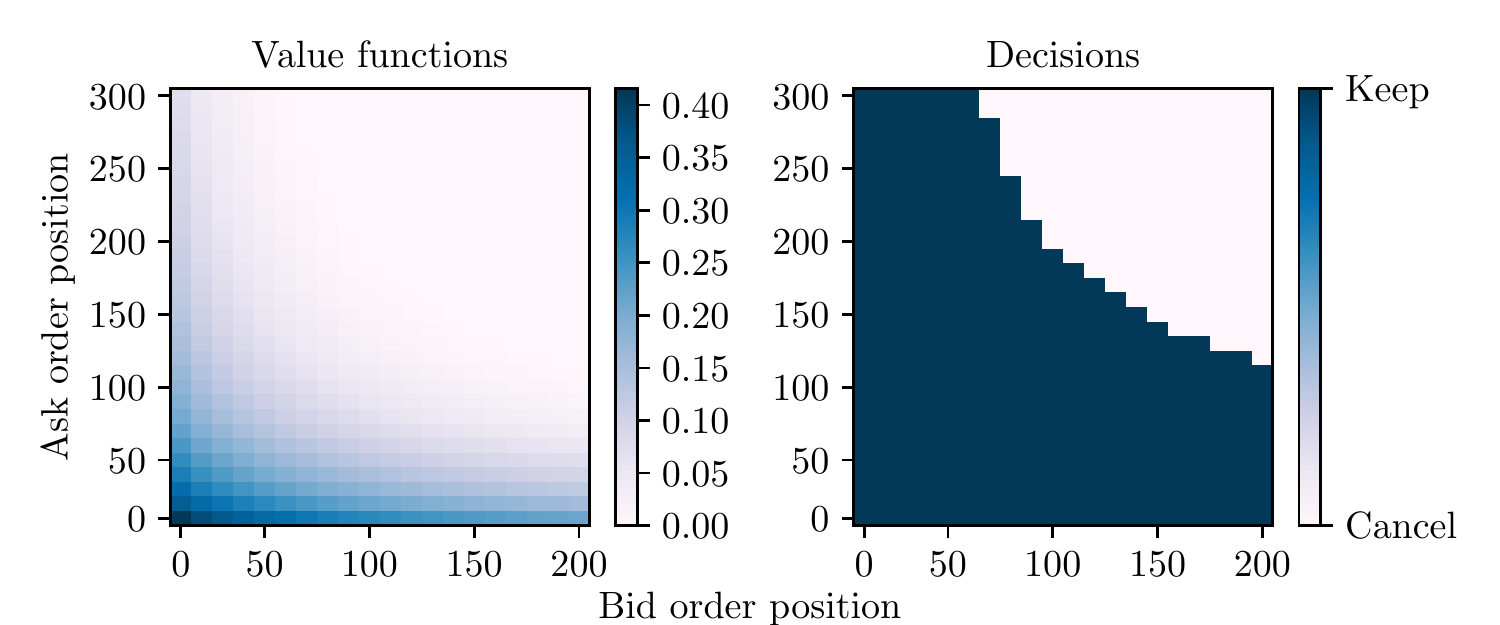}
\caption{State values and decisions when the order book state in $x = (200,300)$}
\label{fig: values_Huang_2030}
\end{figure}

The relative values of different bid and ask positions are also of great importance. Figure \ref{fig: values_Huang_2030} shows the example of the state $x=(200,300)$. As expected, the states become more valuable as the MM's orders get closer to the top of the queues. In addition, the values are not symmetric: the value of $(200, 300, 50, 10)$ is larger than that of $(200,300,10,50)$ because, the longer the queue, the more valuable the priority.

\begin{figure}[ht!]
\center
\includegraphics{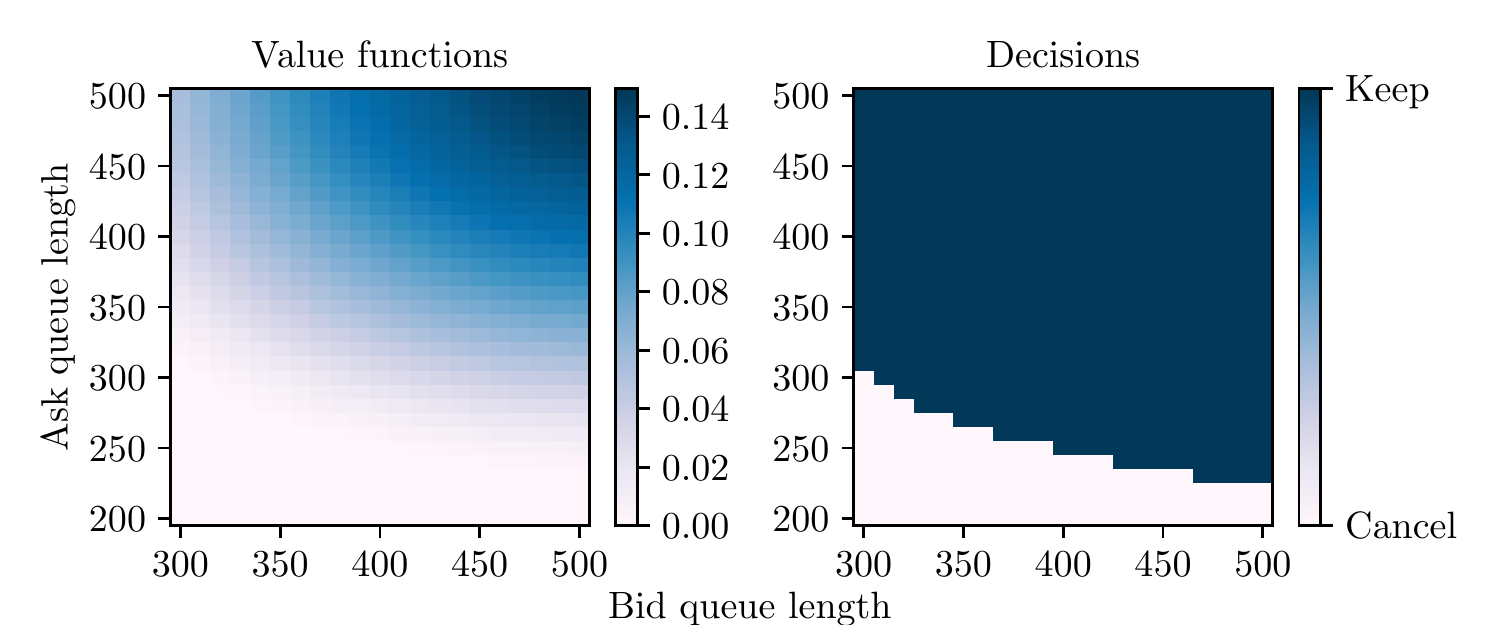}
\caption{State values and decisions when the market maker' order priorities are $(y^B, y^A) = (300,200)$}
\label{fig: values_Huang_FixPos}
\end{figure}

To further illustrate the differences in values and strategies for different states, let us now fix the positions in the queues to be $y^B=300$ and $y^A=200$. Figure \ref{fig: values_Huang_FixPos} is a plot of the value function and optimal decision according to queue lengths.

As expected, the priorities become more valuable when the queues are longer, and the asymmetry also exists. For instance, in this case, and regardless of the bid queue length, the MM should not stay in the order book when the ask queue length is 220 or less.

\begin{figure}[ht!]
\center
\includegraphics{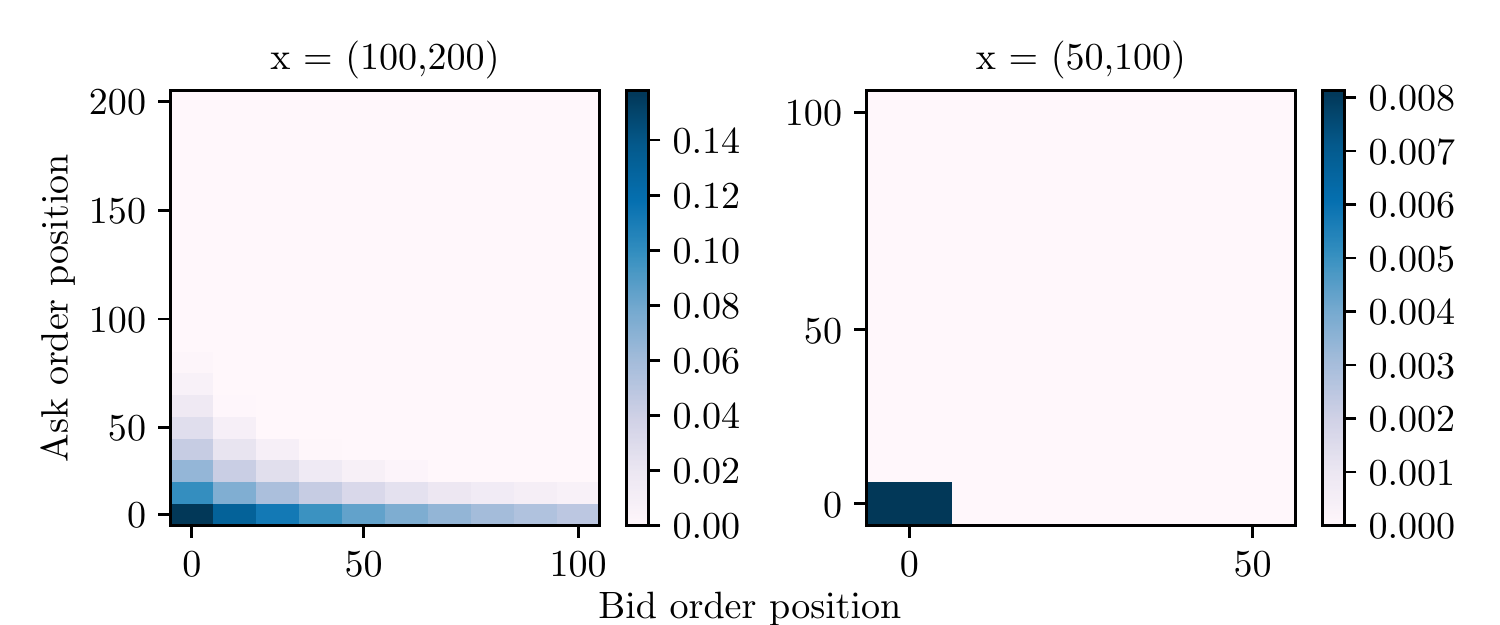}
\caption{State values for short queues}
\label{fig: values_Huang_small}
\end{figure}

Figure \ref{fig: values_Huang_small} shows the value function according to the MM's priority for small queue sizes. When the queue lengths become small, the proportion of market orders compared to that of limit orders increase and the risk of a price move becomes higher. As a consequence, it becomes uninteresting for the MM to stay in the market when the order book is in the state $(50,100)$, except when he is at the top of both queues.

\begin{figure}[ht!]
\center
\includegraphics{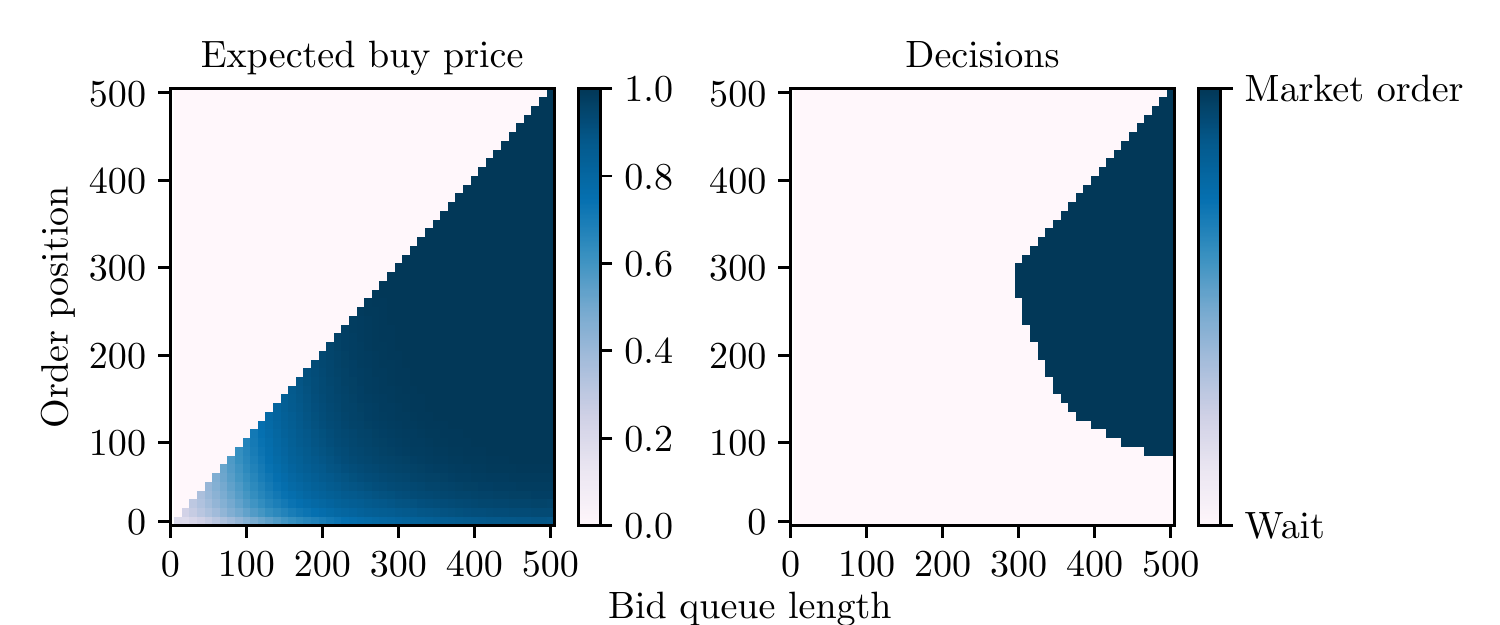}
\caption{Buy-one-lot strategy when $x^A=100$}
\label{fig: values_Huang_BuyOne10}
\end{figure}

\begin{figure}[ht!]
\center
\includegraphics{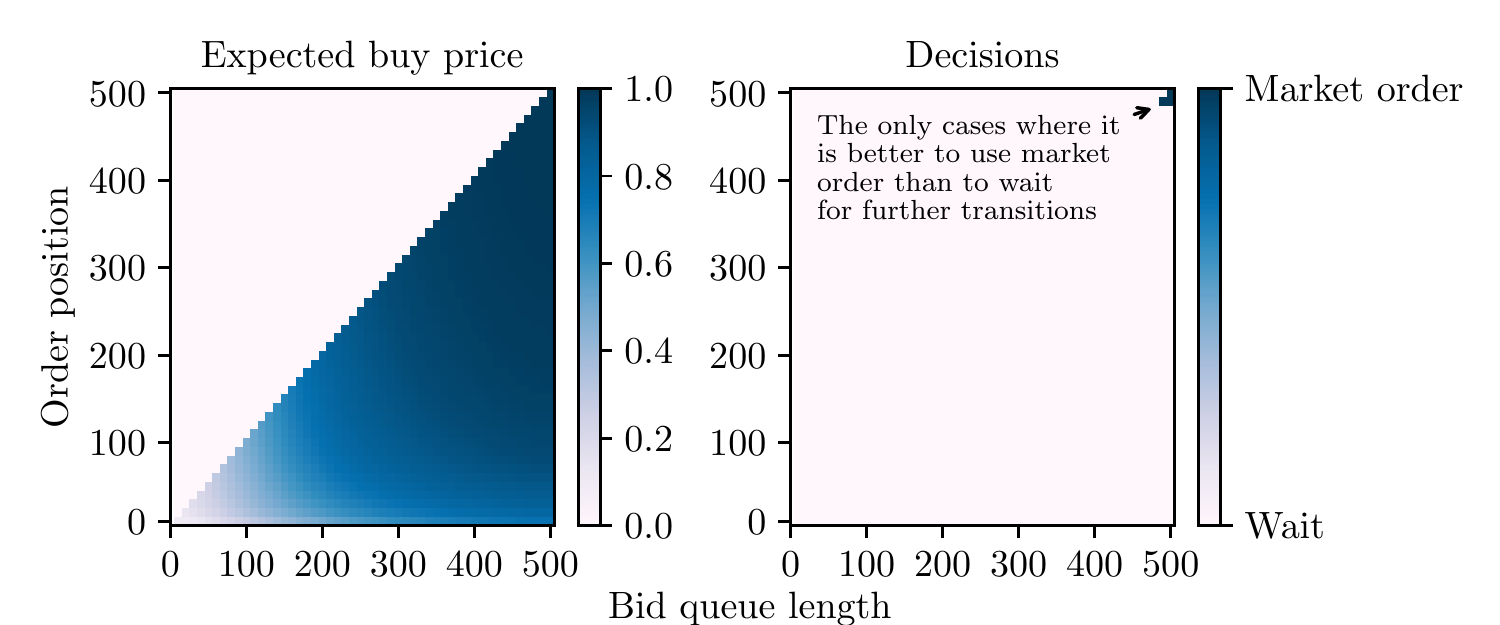}
\caption{Buy one unit strategy when $x^A=140$}
\label{fig: values_Huang_BuyOne14}
\end{figure}

When one order is executed, the strategy becomes a \lq buy-one-lot\rq or \lq sell-one-lot\rq problem. The optimal buy-one-lot strategy after the MM's ask order has been executed is described in Figure \ref{fig: values_Huang_BuyOne10} and Figure \ref{fig: values_Huang_BuyOne14}. Actually, an ask queue length of $x^A=140$ acts as a threshold, beyond which the market maker should always stay in the bid queue and hope for an execution as a limit order. Otherwise, if the ask queue is too short (or the bid queue to long), the price tends to go upwards, and the MM may be better off using a market buy to close his inventory.

\subsubsection{Relaxing the inventory constraints}

The previously defined market making strategies allowed for an inventory of at most 1, a severe restriction in practice.
Relaxing this constraint, one may decide to continuously submit limit bid or ask orders when the previous ones are executed and the state immediately after the execution has value greater than 0, regardless of the inventory. This new strategy violates the initial constraint but however provides some useful information as to whether the profits made by making the spread can cover the inventory risk. This new strategy will be referred to as a \lq locally optimal strategy \rq. As previously seen, the value function for Model 0 is positive for all states, which implies that the MM should follow a naive strategy, continuously submitting orders whatever the state of the LOB. This naive strategy will be used as a benchmark.


Under the assumptions of Model II, Monte Carlo simulation is used to compare the performance of the locally optimal and naive strategies.

For the locally optimal strategy, the simulation runs for an hour of market activity, whereas for the naive strategy, it does for only 10 minutes as the turnover of the naive strategy is much higher. The final inventory is supposed to be closed at the end of the simulation using market orders. 

The daily average P\&L, absolute inventory and turnover - measured in number of contracts - of both strategies are summarized in Table \ref{Tab:simulation}, the standard deviations of the means being given between parentheses. The P\&L of the continuous strategy is much better than that of naive strategy. Its P\&L is significantly positive, with a Z-score of 2.87, whereas the P\&L of the naive strategy is fairly negative, because of the adverse selection effects embedded in Model II. 

\begin{table}[ht!]
\begin{center}
\begin{threeparttable}
\renewcommand{\arraystretch}{1.15}
\caption{Continuous and naive strategy Monte Carlo simulation}
\label{Tab:simulation}

\begin{tabular}{ccc}
\hline
        &continuous	&naive\\ \hline
P\&L	& 1.35 (0.47)& -7.12(0.71) \\
$|$Inv$|$	& 3.33 (0.02)& 10.47(0.09) \\
turnover	& 19.01 (0.06)& 113.46(0.22) \\ \hline
\end{tabular}

\end{threeparttable}
\end{center}
\end{table}

\subsection{Backtesting the optimal strategies}

To better assess the performance of Model II in reproducing a real trading environment, the optimal strategy is backtested using tick-by-tick order book data.

Since building a backtester that can replay historical data in a realistic way is a notably difficult task, we first present our methodology before showing the results. 

\subsubsection{The backtester}

The purpose of a backtesting engine is to reproduce as well as possible the performances of a trading strategy, were it to really be traded in the market. It can produce results that are quite different from those obtained by simulation.

One of the first issues is latency. As pointed out in Figure \ref{fig: spread_switch_dt}, the round-trip latency is clearly reflected in market data. In practice, the incoming data feed and outgoing orders can have different latencies, as they do not necessarily share the same venue (public broadcast reception and private sending). When one reacts to exogeneous information, or information from markets that do not share the same location, these two latencies have to be specified separately. In addition, the latency is not constant, and is typically higher during intense activities because the sequential processing of orders by the matching engine will take more time when orders are piling up. Unfortunately, this variable latency is not measurable, so that for our backtests, a fixed round-trip latency is considered.

Another, very important source of discrepancy between backtests and real market conditions, is of course market impact. Orders in the market are seen by other traders and become a source of information that influences the order flows. For example, a constantly monitored indicator is the market (LOB) imbalance, a clue to short term price movements: an order on the bid side posted by the market maker will actually tend to decrease the probability of execution of said order. However, the market impact of passive limit orders is smaller, and much harder to model, than that of market orders, and we have chosen to ignore it. Such a simplification is realistic when the order size is small compared to the typical queue size. In the case of SX5E futures, the lot size of 10 contracts is very small compared to the average quantity at the first limit, generally around 500 contracts.

When replaying the LOB and trades adding the MM's orders, the key issue is to set some priority rules these fictitious orders in the LOB.

The main question is that of cancellations. Since the exact order flow is not available, it is impossible to know which orders have been completely cancelled. Moreover, modification of an order is allowed with a loss of priority (it can be viewed as simultaneously cancelling an order and resubmitting a new one), so that a decrease in size is not distinguishable from a cancellation. For the sake of simplicity, we decide to randomly choose the position where a cancellation occurs. Since orders with low priorities are more likely to be canceled, a capped exponential law is used.

It is true that there are some ways to improve the identification of canceled orders, by registering the limit orders in a list and matching the quantities, but it is our practical experience that the improvement is marginal.

As for trading rule, a fictitious MM's order will be executed if its position in the queue is within the size of an incoming market order. Immediately after this trade, and before the MM submits another order, the queue length is set to be the same as that observed in the initial LOB data - as if we had increased the trade size to absorb the MM's order.

One important exception is the case when the market order clears one limit: even if the MM's order is at the bottom of the queue, it will be considered to be executed, in accordance with the markedly very high proportion of market orders that actually empty limit in real data. Comparisons with production results show that this choice can influence up to $20\%$ of the turnover for any given strategy and morevoer, that ignoring these executions result in overestimating the backtest performances.

As a conclusion, although the only way to validate a strategy is probably to run it in the market, backtesting engines are always useful, but a lot of care must be taken when designing them and interpreting the results they provide.


\subsubsection{Backtesting market making strategies}

The \lq locally optimal strategy \rq is backtested, as well as a naive strategy (possibly with a threshold).

The backtest is run on the period ranging from July to mid-November, 2016. The MM's order size is fixed to 1 lot (10 contracts).
In the case of the locally optimal strategy, the value function calculated using market data is used to determine whether the MM orders should wait for execution or be cancelled. When one of the orders is executed, another order on the same side is submitted immediately if the value is positive, otherwise, a new order will not be submitted and the existing order on the opposite side is cancelled.

For the naive strategy, bid and ask orders are continuously submitted as soon as the previous one on the same side is executed. When a threshold $q_{min}$ is set, the order stays in the LOB only if the corresponding queue is longer than the threshold.

Two different thresholds of respectively 250 and 400 contracts are studied.


Finally, for practical reasons, we also implement a simple inventory control: whenever the inventory reaches a certain level, new limit orders will no longer be submitted until the inventory falls below the level. In the examples shown here, the maximum inventory is 80, but other values have been tested and do not change the conclusions.


\subsubsection{Results}

\begin{figure}[ht!]
\center
\includegraphics{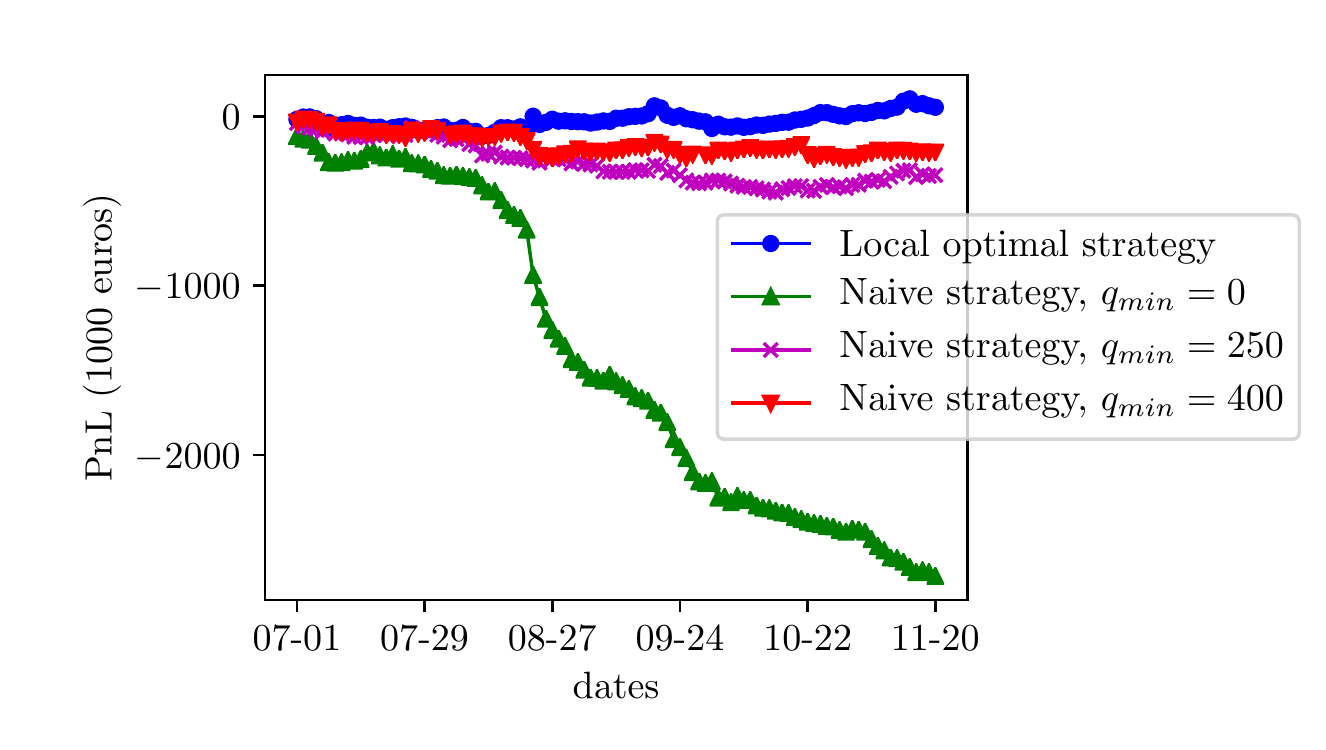}
\caption{Strategies backtested with real data}
\label{fig: backtest_op}
\end{figure}

The backtesting results are presented in Figure \ref{fig: backtest_op}. Table \ref{Tab:backtest} summarizes the daily average P\&L, turnover and profitability of the different strategies. Clearly, the locally optimal strategy is much more profitable than the naive strategies. Without a threshold, the naive strategy provides too much liquidity and the turnover is much higher than with the optimal strategy, making it difficult to compare the P\&Ls. But even with a threshold of 400 (so as to match the turnover of the optimal strategy), the strategy has a decreasing trend and ends up with a negative P\&L.

The locally optimal strategy is the only one that actually ends up positive.  

\begin{table}[ht!]
\begin{center}
\begin{threeparttable}
\renewcommand{\arraystretch}{1.15}
\caption{Average daily statistics for different strategies}
\label{Tab:backtest}

\begin{tabular}{c|c|ccc}
\hline
        &Optimal	&\multicolumn{3}{c}{Naive}\\\hline
         & & $q_{min}=0$ & $q_{min}=250$ & $q_{min}=400$\\ \hline
P$\&$L ($k\euro$) & 0.54 &-26.89 &-3.43 &-2.13 \\
Turn over ($M\euro$) & 45.20 &3430.42 &119.25 &35.83 \\
Profitability (bp) & 0.12 &-0.08 &-0.29 &-0.59 \\
 \hline

\end{tabular}
\end{threeparttable}
\end{center}
\end{table}

%
%

\section{Conclusion}

This paper addresses the calibration of Markovian limit order book models \emph{à la} \cite{Huang:2015}, and their practical usefulness for market making strategies.

First, we show that the size and arrival times of limit orders and cancellations are stable across different queue lengths, whereas market order behave completely differently. Moreover, our analysis shows a strong dependence of a limit establishing order on the nature of the limit removal order. Incorporating these two features allows us to enhance in a significant way the queue-reactive model.

Second, the model is used with the purpose of designing optimal market making strategies. The optimal strategy is compared with the naive strategy, first in a simulation framework where it obviously performs better, but also in a historical backtester, where we also find it to perform much better. This is an indirect proof that the enhanced queue-reactive model may be closer to describing the real market than most Markovian LOB models.

\begin{appendix}
\section{Markov decision processes and optimal strategies}
\label{appendix:MDP}
Let $(X_n)_{n=0}^\infty$ be a Markov chain in discrete time on a countable state space $\mathbb{S}$ with transition matrix $P$. Let $\mathbf{A}$ be a finite set of possible actions. Every action can be classified as either continuation action or termination action. The set of continuation actions is denoted $\mathbf{C}$ and the set of termination actions $\mathbf{T}$. 
$$\mathbf{C}\cup\mathbf{T} = \mathbf{A} \qquad \mathbf{C}\cap\mathbf{T} = \varnothing$$
The Markov chain is terminated when a termination action is selected.

Every action is not available in every state of the chain. Let $A : \mathbb{S}\mapsto 2^{\mathbf{A}}$ be a function associating a non-empty set of actions $A(s)$ to each state $s\in\mathbb{S}$. $2^\mathbf{A}$ is the power set consisting of all subsets of $\mathbf{A}$. The set of continuation actions available in state $s$ is denoted $\mathbf{C}(s) = A(s)\cap \mathbf{C}$ and the set of termination $\mathbf{T}(s) = A(s) \cap \mathbf{T}$. For each $s, s'\in\mathbb{S}$ and $a\in A(s)$ the transition probability from $s$ to $s'$ when selecting action $a$ is denoted $P_{ss'}(a)$.

For every action there are associated values. The value of continuation is denoted $v_C(s,a)$, which can be non-zero only when $a\in\mathbf{C}(s)$. The value of termination is denoted $v_T(s,a)$, it can be non-zero only when $a\in\mathbf{T}(s)$. It is assumed that both $v_C$ and $v_T$ are non-negative and bounded.

A policy $\alpha = (\alpha_0,\alpha_1,\ldots)$ is a sequence of functions: $\alpha_n : \mathbb{S}^{n+1} \mapsto \mathbf{A}$ such that  $\alpha_n(s_0,\ldots,s_n)\in A(s_n)$ for each $n\geq 0$ and $(s_0,\ldots, s_n)\in\mathbb{S}^{n+1}$.

The expected total value starting in $X_0=s$ and following a policy $\alpha$ until termination is denoted by $V(s,\alpha)$. It can be interpreted as the expected payoff of a strategy starting from state $s$ by take the policy $\alpha$. The purpose of Markov decision theory is to analyse optimal policies and optimal expected values. A policy $\alpha_*$ is called optimal if, for all states $s\in\mathbb{S}$ and policies $\alpha$,
$$V(s,\alpha_*)\geq V(s,\alpha)$$
The optimal expected value $V_*$ is defined by
$V_*(s) = \sup_{\alpha}V(s,\alpha)$

If an optimal policy $\alpha_*$ exists, then $V_*(s) = V(s,\alpha_*)$. It is proved in \cite{hult2010algorithmic} that, if all policies terminate in finite time with probability 1, an optimal policy $\alpha_*$ exists and the optimal expected value is the unique solution to a Bellman equation. Furthermore, the optimal policy $\alpha_*$ is stationary, that is to say the policy does not change with time. The optimal values as well as the associated stationary optimal policies can be approached by a recursive algorithm Algorithm 1.

\begin{algorithm}
	\caption{Optimal strategy value approximation}
	\label{algorithm}
	\begin{algorithmic}
		\State \textbf{Input}. Tolerance TOL, transition matrix P, state space $\mathbb{S}$, continuation actions $\mathbf{C}$, termination strategies $\mathbf{T}$,continuation value $v_C$, termination value $v_T$.
		\State \textbf{Output}. Lower bound of optimal value $V_n$ and almost optimal policy $\alpha_n$.
		\While{ d $>$ TOL} 
		\State Put
		\State $$V_n(s) = \max(\max_{a\in\mathbf{C}(s)} v_C(s,a)+\sum_{s'\in\mathbb{S}}P_{ss'}(a)V_{n-1}(s'),\max_{a\in\mathbf{T}(s)}v_T(s,a))$$
		\State and \begin{align*}
			d &= \max_{s\in\mathbb{S}}V_n(s)-V_{n-1}(s) \text{, for } s\in\mathbb{S}\\
			n&=n+1
		\end{align*}
		\EndWhile
		\State Define $\alpha :\mathbb{S}\mapsto \mathbf{C}\cup\mathbf{T}$ as a maximizer to 
		\State $$\max(\max_{a\in\mathbf{C}(s)} v_C(s,a)+\sum_{s'\in\mathbb{S}}P_{ss'}(a)V_{n-1}(s'),\max_{a\in\mathbf{T}(s)}v_T(s,a))$$
	\end{algorithmic}
\end{algorithm}

\subsection{Keep or cancel strategy for buying one unit}

Denote $X_n$ the order book state after $n$ transitions and $X_0$ the initial state. An agent wants to buy one unit at price $j_0<j^A(X_0)$. After each market transition, the agent can choose between keeping the limit order or cancelling it and submit a market buy order at the best ask level $j^A(X_n)$ if the price is lower than the predetermined stop loss price $J>j^A(X_0)$. If $j^A$ reaches $J$ before the agent's order is executed, he cancels the bid order and places a market order at $J$ to fulfil the trade. It is assumed that there are always sufficient limit orders at level $J$.

Denote $Y_n$ the position of the limit order of the agent, and $(O^r_n,q^r_n)$ the last limit removal order type and quantity, where n is the number of event orders from time 0. $S_n = (X_n,Y_n,O^r_n,q^r_n)$ is still a Markov chain in $\mathbb{S}\subset\mathbb{N}^{2K}\times\{0,1,2, \ldots\}\times\{O^M,O^C\}\times\mathbb{N}$ where $Y_n\leq X_n^{j_0}$.

The generator matrix of $S$ is denoted $W = (W_{ss'})$. The jump chain associated with the process is denoted as $S = (S_n)_{n=0}^\infty$. The jump chain is of greater importance in the model. Without ambiguity, we will use $S$ to represent both the continuous process and the jump chain. The transition matrix of the jump chain is denoted $P = P_{ss'}$.

Let $s = (x,y)\in \mathbb{S}$. There are three possible cases:
\begin{itemize}
	\item $y<0$ and $j^A(x)<J$. Then the possible continuation action is $\mathbf{C}(s) = \{0\}$, representing waiting for next market transition. And the possible termination action is $\mathbf{T}(s) = \{-1\}$, representing cancellation of the limit order and submission of market order at $j^A(x)$.
	\item $y<0$ and $j^A(x)=J$. The process terminates as the ask price reaches stop loss price. The limit order is cancelled and a market order is submitted at $j^A(x)=J$, represented by $\mathbf{T}(s) = \{-1\}$. And $\mathbf{C}(s) = \varnothing$.
	\item $y=0$. The process terminates with the execution of the limit order, represented by $\mathbf{T}(s) = \{-2\}$. And $\mathbf{C}(s) = \varnothing$.
	
\end{itemize}
The 1 tick hypothesis is important here for the boundary conditions. Once $J^B=J-2$ and $J^A=J-1$, and the ask is cleared, without the hypothesis we should have terminated the process, except that we have ignored the possibility of the price reversion by a new ask order, so that the execution probability of the market maker's bid order is underestimated.

The expected value (cost), interpreted as the expected saving with respect to stop loss price, is given by 

\begin{equation}
V_\infty(s,\alpha)=\left\{
	\begin{array}{lll}
	\sum_{s'\in\mathbb{S}} P_{ss'}V_\infty(s',\alpha) &,& \alpha(s)=0\\
	\pi^{J}-\pi^{j^A(x)} &,& \alpha(s)=-1\\
	\pi^{J}-\pi^{j_0} &,& \alpha(s)=-2
	\end{array}\right.
\end{equation}

The waiting value is zero. The value function could then be approximated by Algorithm  with the iteration
\begin{align*}
	V_{n+1}(s) &= \max(\max_{a\in\mathbf{C}(s)} \sum_{s'\in\mathbb{S}}P_{ss'}(a)V_{n}(s'),\max_{a\in\mathbf{T}(s)}v_T(s,a))\\
	&=\left\{
	\begin{array}{lll}
	\max(\sum_{s'\in\mathbb{S}} P_{ss'}V_{n}(s'),\pi^{j^A}(s)) &,& \text{for } y>0, j^A<J\\
	\pi^J-\pi^J=0 &,& \text{for } y>0, j^A=J\\
	\pi^{J}-\pi^{j_0} &,& \text{for } y=0
	\end{array}
	\right.
\end{align*}

\subsection{Market making (Making the spread)}
\label{section3:MM}

The extended Markov chain here is defined as $(X_n,Y_n^0,Y_n^1,j_n^0,j_n^1,O^r_n,q^r_n)$, where $Y_n^0$($Y_n^1$) is the positions of the market maker's bid(ask) order in the bid(ask) price level $j_n^0$ ($j_n^1$). We have both $Y_n^0$ and $Y_n^1$ are non-increasing, and
$$X_n^{j_n^0}\geq Y_n^0\geq 0 \qquad X_n^{j_n^1}\geq Y_n^1\geq 0$$

The market maker predetermines a best buy level $J^{B^0}<j^A(X_0)$, a worst level $J^{B^1}>j^A(X_0)$, a best sell level $J^{A^1}>j^B(X_0)$ and a worst sell level $J^{A^0}<j^B(X_0)$. As in the buy one strategy, the order is cancelled and executed at the stop loss price if the corresponding best limit price reaches the worst price level. And it is assumed that the execution at the stop loss price is always available. The state space is defined on $\mathbb{S}\subset \mathbb{N}^d\times\{0,1,2,\ldots\}\times\{0,1,2,\ldots\}\times\{J^{B^0},\ldots,J^{B^1}-1\}\times\{J^{A^0}+1,\ldots,J^{A^1}\}\times\{O^M,O^C\}\times\mathbb{N}$.

The possible actions in this strategy are:
\begin{itemize}
 \item The market maker can choose to wait for next market transition and cancel both orders before any of the orders is executed
 \item When one of the orders has been executed, the market maker has one order on the opposite side waiting for execution. The market maker follows a buy(sell)-one-unit strategy.
\end{itemize}

Let $V_\infty^B(x,y,j,o^r,q^r)$ denote the optimal(minimal) expected buy price (cost rather than value) in state $(x,y,j,o^r,q^r)$ for buying one unit, with best buy level $J^{B^0}$ and worst level $J^{B^1}$. Similarly, $V_\infty^A(x,y,j)$ denotes the optimal (maximal) expected sell price in state $(x,y,j,o^r,q^r)$ for selling one unit, with best sell level $J^{A^1}$ and worst sell level $J^{A^0}$. The optimal expected value is then given by
$$V_\infty(s)
\left\{
\begin{array}{lll}
\max(\sum_{s'\in\mathbb{S}}P_{ss'}V_\infty(s'),0) &, & \text{for } y^0>0, y^1>0\\
\pi^{j^1}-V_\infty^B(x,y^0,j^0,o^r,q^r) &, & \text{for } y^0>0, y^1=0\\
V_\infty^A(x,y^1,j^1)-\pi^{j^0,o^r,q^r} &, & \text{for } y^0=0, y^1>0
\end{array}
\right.
$$

An extended version is also possible, to take the change of limit price into consideration. Under this strategy, the available actions for the market maker are:
\begin{itemize}
 \item Before any of the orders is executed, the market maker can choose from waiting for next transition, cancel both orders or cancel either order and resubmit at new levels $k^0$ et $k^1$.
 \item When one of the orders have been processed, the outstanding limit order is proceeded according to the buy(sell)-one-unit strategy. And the price level is also renewable after each market transition.
\end{itemize}

In this strategy, the optimal expected value is determined by
$$V_\infty(s)
\left\{
\begin{array}{lll}
\max(\sum_{s'\in\mathbb{S}}P_{ss'}V_\infty(s'), \max V_\infty(s_{k^0k^1}) ,0) &, & \text{for } y^0>0, y^1>0\\
\pi^{j^1}-V_\infty^B(x,y^0,j^0,o^r,q^r) &, & \text{for } y^0>0, y^1=0\\
V_\infty^A(x,y^1,j^1,o^r,q^r)-\pi^{j^0} &, & \text{for } y^0=0, y^1>0
\end{array}
\right.
$$
where $s_{k^0k^1}$ describes the cancel and resubmit of one limit order.

\end{appendix}

\bibliographystyle{apalike}
\bibliography{MM_Markov}
\end{document}